\documentclass[prb,aps,onecolumn,showpacs,floats,a4paper,times,12pt]{revtex4}
\usepackage{epsfig,bm,epsf,graphics}
\usepackage{xcolor}

\begin{document}
%\begin{frontmatter}
\title{Dzyaloshinskii-Moriya Interaction in Magneto-Ferroelectric Superlattices: Spin Waves and Skyrmions}
\author{I. F. Sharafullin $^{a,b}$, M. Kh. Kharrasov $^{b}$, H. T. Diep $^{a}$}
\address{$^{a}$ Laboratoire de Physique Th\'eorique et Mod\'elisation
Universit\'e de Cergy-Pontoise, CNRS, UMR 8089, 2 Avenue Adolphe
Chauvin, 95302 Cergy-Pontoise, Cedex, France.\\
 $^{b}$ Bashkir State University, 32, Validy str, 450076, Ufa, Russia.}

\date{\today}

\begin{abstract}
We study in this paper effects of Dzyaloshinskii-Moriya (DM)
magnetoelectric coupling between ferroelectric and magnetic layers
in a superlattice formed by alternate magnetic and ferroelectric films.
Magnetic
films are films of simple cubic lattice with Heisenberg
spins interacting with each other via an exchange $J$ and a DM interaction with the ferroelectric interface. Electrical polarizations of $\pm{1}$ are assigned
at simple cubic lattice sites in the ferroelectric films.
We determine the ground-state (GS) spin configuration in the magnetic film. In zero field, the GS is periodically non collinear and in an applied field $\mathbf H$ perpendicular to the layers, it shows the existence of skyrmions at the interface. Using the Green's function method we study the spin waves (SW) excited in a monolayer and also in a bilayer sandwiched between ferroelectric films, in zero field. We show that the DM interaction strongly affects the long-wave length SW mode. We calculate also the magnetization at low temperature $T$.  We use next Monte Carlo simulations to calculate various physical quantities at finite temperatures such as the critical temperature, the layer magnetization and
the layer polarization, as functions of the magnetoelectric DM coupling and the applied magnetic field.
Phase transition to the disordered phase is studied in detail.
\end{abstract}


\maketitle

\section{Introduction}\label{sect-intro}

\begin{figure}[h]
\begin{center}
\includegraphics[scale=0.53]{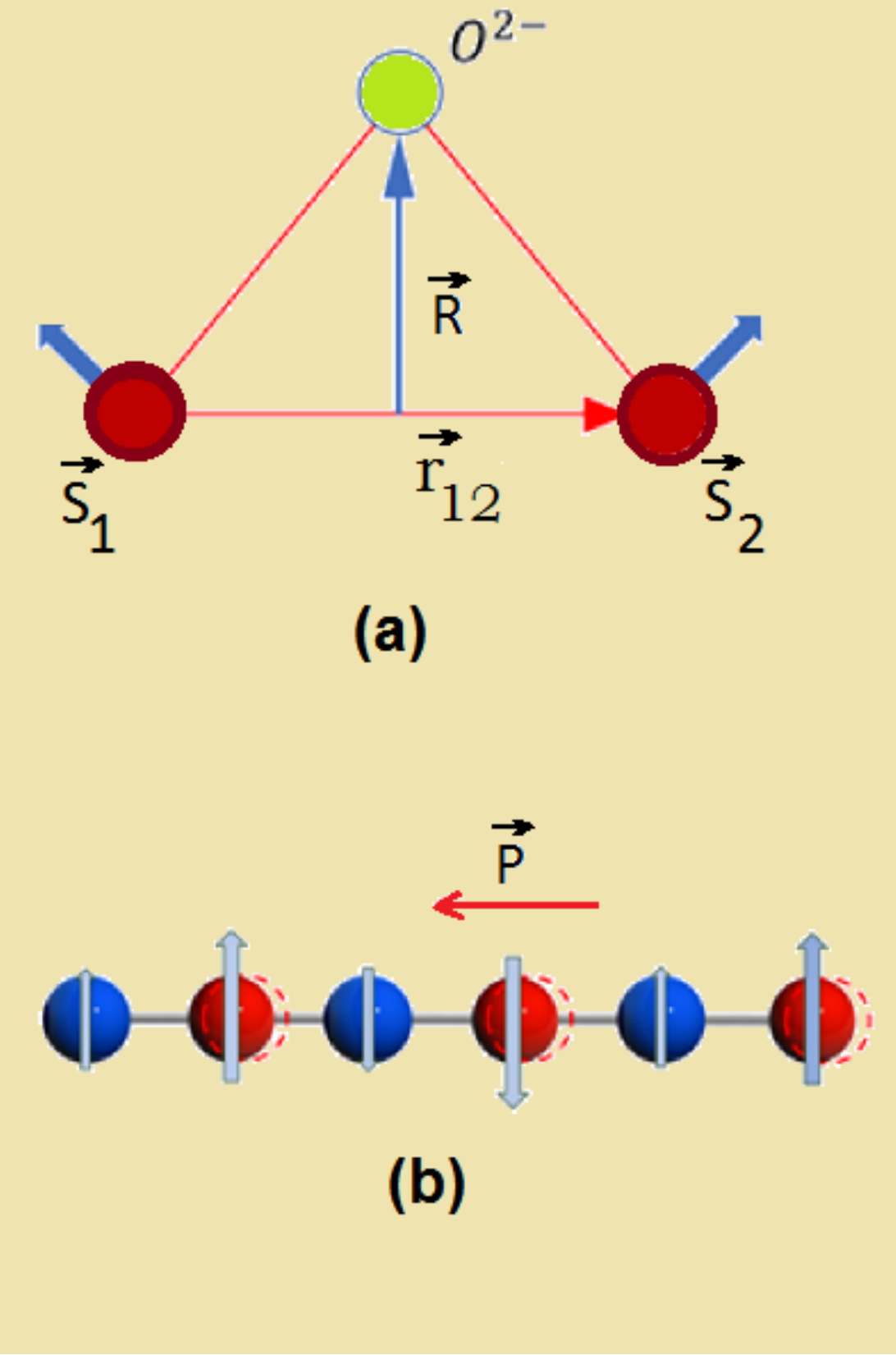}
\caption{(a) Schema of Dzyaloshinskii-Moriya interaction  (b) microscopic mechanisms of creation of electric polarization $\vec P$ due to displacements of atoms (red) in the region of inhomogeneous distribution of magnetizations.}
\label{ref-fig1a}
\end{center}
\vspace{-10pt}
\end{figure}

Non-uniform spin structures, which are quite interesting by
themselves, became the subject of close attention after the
discovery of electrical polarization in some of them
\cite{dong2011microscopic}. The existence of polarization is
possible due to the inhomogeneous magnetoelectric effect, namely
that electrical polarization can occur in the region of magnetic
inhomogeneity. It is known that the electric polarization vector is
transformed in the same way as the combination of the magnetization
vector and the gradient of the magnetization vector, meaning that
these values can be related by the proportionality relation. In Ref.
\onlinecite{mostovoy2006ferroelectricity} it was found that in a crystal
with cubic symmetry the relationship between electrical polarization
and inhomogeneous distribution of the magnetization vector has the
following form
\begin{equation}\label{1}
\mathbf{P}=\gamma\chi_{e}(\mathbf{M}\cdot(\bigtriangledown\cdot\mathbf{M})
-(\mathbf{M}\cdot\bigtriangledown)\cdot\mathbf{M})
\end{equation}
here $\gamma$ is the magnetoelectric coefficient, and $\chi_{e}$ the permittivity.
%In particular, for the family of orthorhombic manganites
%$RMn_{2}O_{5}, R=\left(Eu, Gd, Er,Y \right) $, the magnetoelectric contribution
%to the free energy has the form
%$\alpha \cdot P_{y}(A^2-G^2)$, where $\alpha$ is the magnetoelectric interaction constant,
% $\mathbf{P}$ is the vector of electric polarization,
% $\mathbf{A}=(\mathbf{S_{1}}-\mathbf{S_{2}})-(\mathbf{S_{3}}-\mathbf{S_{4}})$ and $\mathbf{G}=(\mathbf{S_{1}}-\mathbf{S_{2}})+(\mathbf{S_{3}}-\mathbf{S_{4}})$ - are antiferromagnetic
%order parameters for four spins of manganese ions Mn$^{3+}$. 
In non collinear structures, the microscopic mechanism of the coupling
of polarization and the relative orientation of the magnetization
vectors is based on the interaction of Dzyaloshinskii-Moriya
\cite{katsura2005spin,sergienko2006role,cheong2007multiferroics}.
The corresponding term in the Hamiltonian is:
\begin{equation}\label{2}
H_{DM}=\mathbf {D}_{i,j}\cdot\mathbf {S_{i}}\times\mathbf {S_{j}}
\end{equation}
where $\mathbf {S_{i}}$ is the spin of the i-th magnetic ion, and
$\mathbf{D}_{i,j}$ is the Dzyaloshinskii-Moriya vector. The vector $\mathbf{D}_{i,j}$
is proportional to the vector product $\mathbf{R}\times\mathbf{r_{i,j}}$
of the vector $\mathbf{R}$ which specifies the displacement of the
ligand (for example, oxygen) and the unit vector $\mathbf{r_{i,j}}$
along the axis connecting the magnetic ions $i$ and $j$ (see Fig.
\ref{ref-fig1a}a). We write

\begin{equation}\label{3}
\mathbf D_{i,j} \propto \ \mathbf {R}\times\mathbf {r_{i,j}}
\end{equation}

%Fig1
Thus, the Dzyaloshinskii-Moriya interaction connects the angle
between the spins and the magnitude of the displacement of
non-magnetic ions. In some micromagnetic structures all ligands are
shifted in one direction, which leads to the appearance of
macroscopic electrical polarization (see Fig. \ref{ref-fig1a}b). By nature, this interaction is
a relativistic amendment to the indirect exchange interaction, and
is relatively weak \cite{pyatakov2012spin}. In the case of
magnetically ordered matter, the contribution of the
Dzyaloshinskii-Moriya interaction to the free energy can be
represented as Lifshitz antisymmetric invariants containing spatial
derivatives of the magnetization vector. In analogy, the vortex
magnetic configuration can be stable via Skyrme mechanism
\cite{bogdanov1989thermodynamically}. Skyrmions were theoretically
predicted more than twenty years ago as stable micromagnetic
structures \cite{bogdanov1994thermodynamically}. The idea came from
nuclear physics, where the elementary particles were represented as
vortex configurations of continuous fields. The stability of such
configurations was provided by the "Skyrme mechanism" - the
components in Lagrangians containing antisymmetric combinations of
spatial derivatives of field components \cite{skyrme1962unified}.
For a long time skyrmions have been the subject only of theoretical
studies. In particular, it was shown that such structures can exist
in antiferromagnets \cite{bogdanov2002magnetic} and in magnetic
metals \cite{rossler2006spontaneous}. In the latter case, the model
included the possibility of changing the magnitude of the
magnetization vector and spontaneous emergence of the skyrmion
lattice without the application of external magnetic field. A
necessary condition for the existence of skyrmions in bulk samples
was the absence of an inverse transformation in the crystal magnetic
symmetry group.  Diep et al. \cite{diep2018skyrmion} have
studied a crystal of skyrmions generated on a square lattice using a
ferromagnetic exchange interaction and a Dzyaloshinskii-Moriya
interaction between nearest-neighbors under an external magnetic
field. They have shown that the skyrmion crystal has a hexagonal
structure which is shown to be stable up to a temperature $T_c$
where a transition to the paramagnetic phase occurs and the dynamics
of the skyrmions at $T < T_c$ follows a stretched exponential law.
In Ref. \onlinecite{rossler2006spontaneous} it was shown that the most
extensive class of candidates for the detection of skyrmions includes the
surfaces and interfaces of magnetic materials, where the geometry of
the material breaks the central symmetry and, therefore, can lead to
the appearance of chiral interactions similar to the
Dzyaloshinskii-Moriya  interaction. In addition, skyrmions are
two-dimensional solitons, the stability of which is provided by the
local competition of short-range interactions exchange and
Dzyaloshinskii-Moriya interactions \cite{diep2018skyrmion,kiselev2011ns}. The idea of
using skyrmions in memory devices nowadays is reduced to the
information encoding using the presence or absence of a skyrmion in
certain area of the material. A numerical simulation of the creation
and displacement of skyrmions in thin films was carried out in Ref.
\onlinecite{sampaio2013nucleation} using a spin-polarized current. The
advantage of skyrmions with respect to the domain boundaries in such
magnetic memory circuits (e.g. racetrack memory, see
Ref. \onlinecite{parkin2008magnetic}) is the relatively low magnitude of the
currents required to move the skyrmions along the "track". For the
first time, skyrmions were experimentally detected in the $MnSi$
helimagnet \cite{muhlbauer2009skyrmion}. Below the Curie temperature
in $MnSi$ spins are aligned in helicoidal or conical structure (the
field was applied along the $[100]$ axis), depending on the
magnitude of the applied magnetic field. Similar experimental
results were obtained for the compound $Fe_{1-x}Co_{x}Si, x = 0.2$
\cite{munzer2010skyrmion}. Note here that properties of a
helimagnetic thin film with quantum Heisenberg spin model by using
the Green's function method was investigated in Ref.
\onlinecite{PhysRevB.91.014436}. Surface spin configuration is calculated
by minimizing the spin interaction energy. The
transition temperature is shown to depend strongly on the
helical angle. Results are in agreement with existing experimental
observations on the stability of helical structure in thin films and
on the insensitivity of the transition temperature with the film
thickness.The investigation of $Fe_{0.5}Co_{0.5}Si$ made it possible
to take the next important step in the study of skyrmions - to
directly observe them using Lorentz electron microscopy
\cite{yu2010real}. The sample was a thin film, magnetic structure of
which can be considered two-dimensional: the spatial period of the
helicoid (90 nm) exceeded the film thickness, therefore its wave
vector laid in the film plane. The magnetic field was applied
perpendicular to the film, resulting in suppression of helix and the
appearance of the skyrmions lattice.
%(see Fig. \ref{ref-fig1b})
%
%%Fig2
%\begin{figure}[h]
%\begin{center}
%\includegraphics[scale=0.53]{fig1b.pdf}
%\caption{Skyrmions in $Fe_{0.5}Co_{0.5}Si$ \cite{yu2010real}: a -
%phase diagram. Numbers mark dots, corresponding to the tie-ins at
%the top of the figure; b - image of the skyrmion lattice obtained
%using Lorentz electron microscopy}\label{ref-fig1b}
%\end{center}
%\vspace{-10pt}
%\end{figure}
%In addition, the authors of Ref. \onlinecite{yu2010real} took pictures of the
%magnetic structure in the transition regions of the phase diagram:
%one can observe  how the helicoid turns into the
%skyrmion lattice, and one can see the separate
%skyrmions in the process of transition between the grid of skyrmions
%and the phase of uniform magnetization in the strong magnetic field.
%In the case of a thin film skyrmions lattice was stable over a wider
%range of temperature and magnetic field. 
The dependence of the
stability of the skyrmion lattice on the sample thickness was
studied in more detail in Ref. \onlinecite{yu2011near}. A wedge-shaped $FeGe$
sample was created, whose thickness varied from 15 nm to hundreds of
nanometers (with a helicoid period of about 70 nm). Studies have
confirmed that the thinner was the film, the greater was the
"stability region" of skyrmions. Skyrmions as the most compact
isolated micromagnetic objects are of great practical interest as
memory elements  \cite{kiselev2011ns}. The stability of skyrmions \cite{diep2018skyrmion} can make the memory on their basis non-volatile, and low control
currents will reduce the cost of rewriting compared to similar
technologies based on domain boundaries. In Refs.
\onlinecite{seki2012observation,seki2012magnetoelectric} magnetic and
electrical properties of the skyrmion lattice were studied in the
multiferroic $Cu_{2}OSeO_{3}$. It has been shown that that energy
consumption can be minimized by using the electric field to control
the micromagnetic structures. It is worth noting that the
multiferroics $BaFe_{12-x-0.05}Sc_{x}Mg_{0.05}O_{19}$ may also have
a skyrmion structure \cite{yu2012magnetic,rosch2012extra}. The
manipulations with skyrmions were first demonstrated in the diatomic
$PdFe$ layer on the iridium substrate, and the importance of this
achievement for the technology of information storing is difficult
to overestimate: it makes possible to write and read the individual
skyrmions using a spin-polarized tunneling current
\cite{romming2013writing}. The idea was to apply the magnetic field
to the region of the phase diagram corresponding to the intermediate
state between the skyrmion lattice and the uniformly magnetized
ferromagnetic state. Then, using a needle of a tunneling microscope,
a spin-polarized current was passed through various points of the
sample, which led to the appearance of skyrmions in the desired
positions. In Ref. \onlinecite{pyatakov2011magnetically}, the possibility of
the nucleation of skyrmions by the electric field by means of an
inhomogeneous magnetoelectric effect was established. The required
electric field strength can be estimated in order of magnitude as
$10^{6} B/cm$, which lies in the range of experimentally achievable
values. It is shown that the direction of the electric field
determines the chirality of the micromagnetic structure. Recent
studies are focused on the interface-induced skyrmions. Therefore,
the superstructures naturally lead to the interaction of skyrmions
on different interfaces, which has unique dynamics compared to the
interaction of the same-interface skyrmions. In Ref.
\onlinecite{koshibae2017theory}, a theoretical study of two skyrmions on
two-layer systems was carried using micromagnetic modeling, as well
as an analysis based on the Thiele equation, which revealed a
reaction between them, such as the collision and a bound state
formation. The dynamics sensitively depends on the sign of DM
interaction, i.e. the helicity, and the skyrmion numbers of  two
skyrmions, which are well described by the Thiele equation. In
addition, the colossal spin-transfer-torque effect of bound skyrmion
pair on antiferromagnetically coupled bilayer systems was
discovered. In Ref. \onlinecite{martinez2016topological} the study of the
Thiele equation was carried for current-induced motion in a skyrmion
lattice through two soluble models of the pinning potential.

We consider in this paper a superlattice composed of alternate magnetic films and ferroelectric films.  The aim of this paper is to propose a new model for the coupling between the magnetic film and the ferroelectric film by introducing a DM-like interaction.  It turns out that this interface coupling gives rise to non collinear spin configurations in zero applied magnetic field and to skyrmions in a field $\mathbf H$ applied perpendicularly to the films.  Using the Green's function method, we study spin-wave excitations in zero field of a monolayer and a bilayer. We find that the DM interaction affects strongly the long wave-length mode. Monte Carlo simulations are carried out to study the phase transition of the superlattice as functions of the interface coupling strength.

The paper is organized as follows. Section \ref{MGS} is devoted to the description of our model and the determination of the ground-state spin configuration with and without applied magnetic field. In section \ref{Green} we show the results of the Green's function technique in zero field for a monolayer and a bilayer. Section \ref{MC} shows the results obtained by Monte Carlo simulations for the phase transition in the system as a function of the interface DM coupling. Concluding remarks are given in section \ref{Concl}.

\section{Model and ground state}\label{MGS}
\subsection{Model}
Consider a superlattice composed of alternate magnetic and ferroelectric films (see Fig. \ref{ref-fig1b}). The Hamiltonian of this multiferroic superlattice is expressed as:

\begin{equation}\label{eq-ham-sysm-1}
{\cal H}=H_{m}+H_{f}+H_{mf}
\end{equation}
where $H_m$ and $H_f$ are the Hamiltonians of the ferromagnetic and ferroelectric subsystems, respectively, while $H_mf$ is the Hamiltonian of magnetoelectric interaction at the interface between two adjacent films.

We describe the Hamiltonian of the magnetic film with the Heisenberg spin model on a cubic lattice:

\begin{equation}\label{eq-ham-sysm-2}
H_{m}=-\sum_{i,j}{J^{m}_{ij}
\mathbf {S}_{i}\cdot \mathbf {S}_{j}}-\sum_{i}\mathbf {H}\cdot \mathbf{S}_{i}
\end{equation}
where $\mathbf S_{i}$  is the spin on the i-th site, $\mathbf {H}$  is the external magnetic field,  $J_{ij}^{m}>0$ the ferromagnetic interaction parameter between a spin and its nearest neighbors (NN) and the sum is taken over NN spin pairs. We consider $J_{ij}^{m}>0$ to be the same, namely $J^m$, for spins everywhere in the magnetic film.
The external magnetic field $\mathbf {H}$ is applied along the $z$-axis which is perpendicular to the plane of the layers.
The interaction of the spins at the interface will be given below.

For the ferroelectric film, we suppose for simplicity that electric polarizations are Ising-like vectors of magnitude 1,  pointing in the $\pm z$ direction.  The Hamiltonian is given by
\begin{equation}\label{eq-ham-sysm-3}
H_{f}=-\sum_{i,j}{J_{ij}^{f} {\mathbf P}_{i}\cdot
{\mathbf P}_{j}}-\sum_{i}E^{z}P^{z}_{i}
\end{equation}
where $\mathbf P_{i}$ is the polarization on the i-th lattice site, $J_{ij}^{f}>0$ the interaction parameter between NN and the sum is taken over NN sites. Similar to the ferromagnetic subsystem we will take the same $J_{ij}^{f}=J^f$ for all ferroelectric sites. We apply the external electric field $\mathbf {E}$ along the $z$-axis.

We suppose the following Hamiltonian for the magnetoelectric interaction at the interface
\begin{equation}
H_{mf}= \sum_{i,j,k}J_{ijk}^{mf}\ \mathbf D_{i,j}\cdot \left [\mathbf {S_{i}}\times \mathbf {{S_{j}}} \right ]\label{Hmf0}
\end{equation}
In this expression $J_{ijk}^{mf}\ \mathbf D_{i,j}$ plays the role of the DM vector which is  perpendicular to the $xy$ plane. Using Eqs. (\ref{2})-(\ref{3}), one has
\begin{eqnarray}
\mathbf D_{i,j}&=&\mathbf R \times \mathbf r_{i,j}\nonumber\\
\mathbf D_{j,i}&=&\mathbf R \times \mathbf r_{j,i}=-\mathbf D_{i,j}
\end{eqnarray}

Now, let us define  for our model
\begin{equation}
J_{ijk}^{mf}=J_{i,j}^{mf}P_k
\end{equation}
which is the DM interaction parameter
between the electric polarization $\mathbf P_{k}$ at the interface
ferroelectric layer and the two NN spins
$\mathbf {S_{i}}$ and $\mathbf {S_{j}}$ belonging to the interface ferromagnetic layer.  Hereafter, we suppose $J_{i,j}^{mf}=J^{mf}$ independent of $(i,j)$.  Selecting $\mathbf R$ in the $xy$ plane perpendicular to $\mathbf r_{i,j}$  (see Fig. \ref{ref-fig1a}) we can write $\mathbf R \times \mathbf r_{i,j}=a\mathbf z\ e_{i,j}$ where $e_{i,j}=-e_{j,i}=1$, $a$ is a constant and $\mathbf z$ the unit vector on the $z$ axis.

It is worth at this stage to specify the nature of the DM interaction to avoid a confusion often seen in the literature. The term $\left [\mathbf S_{i}\times \mathbf S_{j} \right ]$ changes its sign with the permutation of $i$ and $j$, but the whole DM interaction defined in Eq. (\ref{2}) does not change its sign because $D_{i,j}$ changes its sign with the permutation as seen in Eq. (\ref{3}).  Note that if the whole DM interaction is antisymmetric then when we perform the lattice sum, nothing of the DM interaction remains in the Hamiltonian.  This explains why we need the coefficient $e_{i,j}$ introduced above and present in Eq. (\ref{Hmf}).

 We collect all these definitions we write $H_{mf}$ in a simple form

\begin{eqnarray}
H_{mf}&=& \sum_{i,j,k}J^{mf}\ P_k\ (\mathbf R \times \mathbf r_{i,j})\cdot \left [\mathbf {S_{i}}\times \mathbf {{S_{j}}} \right ]\nonumber\\
&=& \sum_{i,j,k}J^{mf}\ P_k \ e_{i,j}\mathbf z \cdot \left [\mathbf {S_{i}}\times \mathbf {{S_{j}}} \right ]\nonumber\\
&=& \sum_{i,j,k}J^{mf}\  e_{i,j}\ \mathbf P_k \cdot \left [\mathbf {S_{i}}\times \mathbf {{S_{j}}} \right ]\label{Hmf}\label{Hmf}
\end{eqnarray}
where the constant $a$ is absorbed in $J^{mf}$.

\begin{figure}[h]
\vspace{-10pt}
\begin{center}
\includegraphics[scale=0.50]{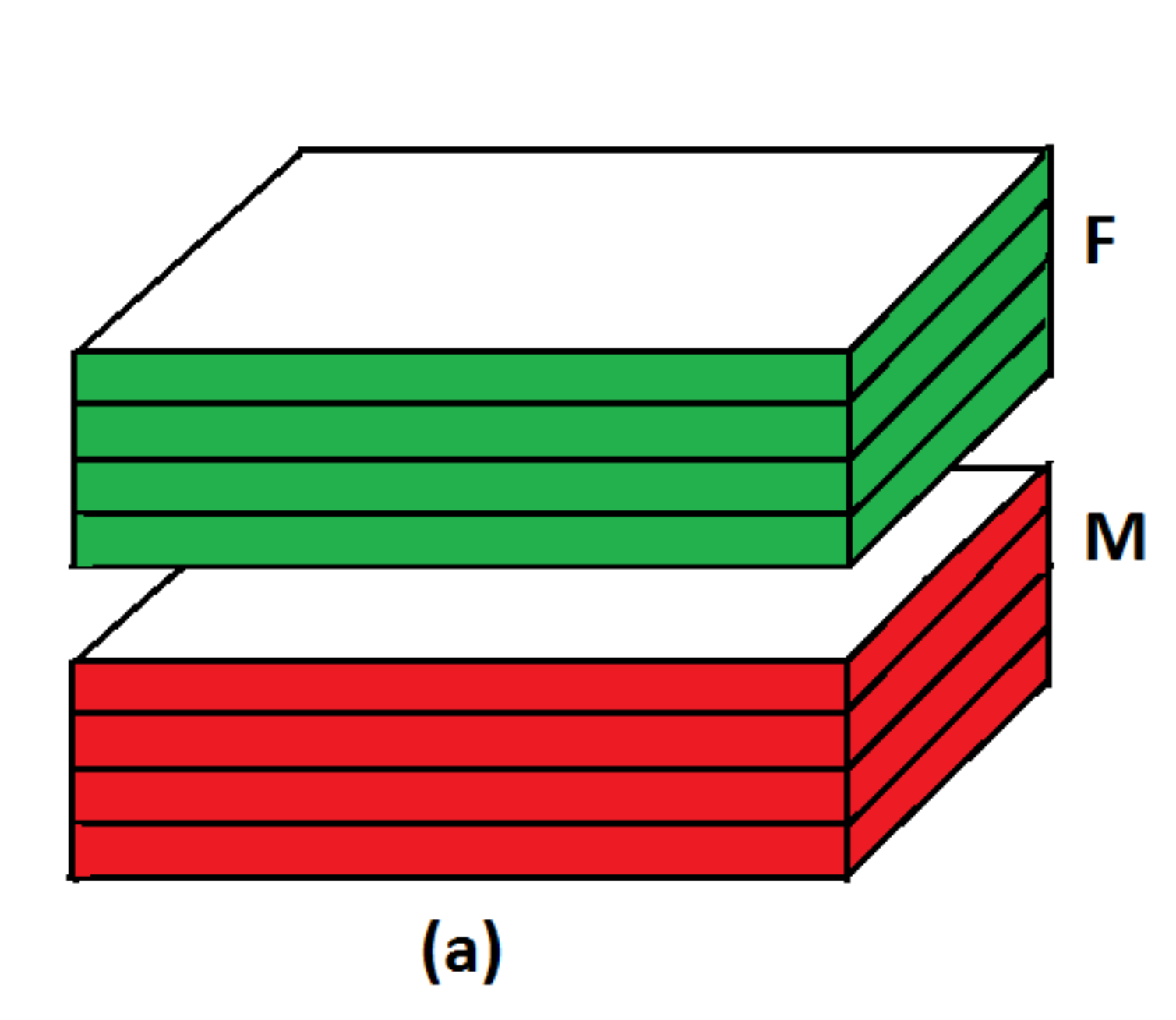}
\includegraphics[scale=0.50]{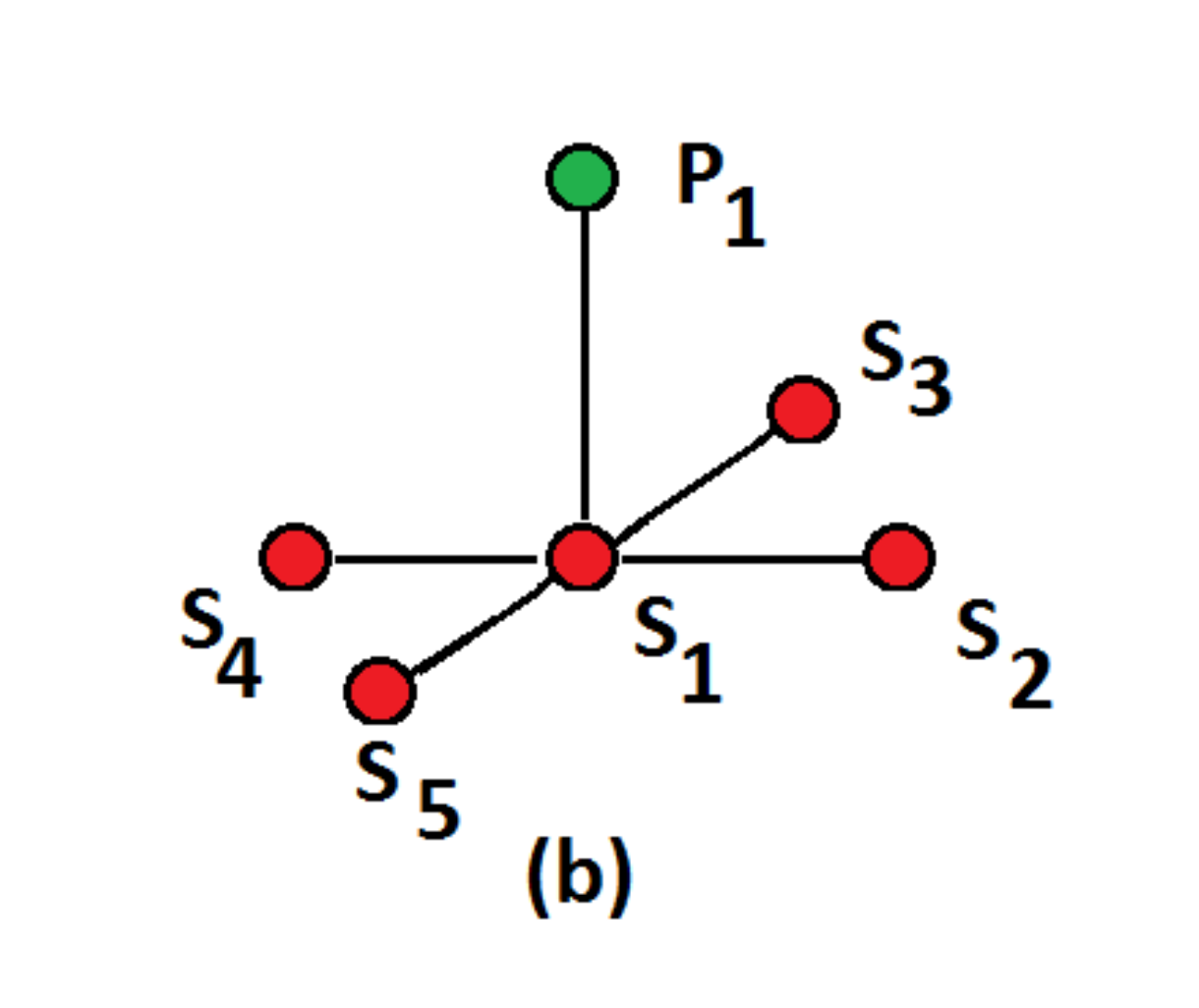}
\end{center}
\vspace{10pt} \caption{(a) The superlattice composed of alternately a ferroelectric layer indicated by F and a magnetic layer indicated by M; (b) A polarization $P_1$ at the interface interacts with 5 spins in the magnetic layer. See text for expression. } \label{Model} \vspace{5pt}
\end{figure}

The superlattice and the interface interaction are shown in Fig. \ref{Model}. A polarization at the interface interact with 5 spins on the magnetic layer according to Eq. (\ref{Hmf}), for example (see Fig. \ref{Model}b):
\begin{eqnarray}
&&J^{mf} \mathbf P_{1} \cdot  [ e_{1,2}  (\mathbf {S_{1}}\times \mathbf {{S_{2}}})
+e_{1,3} (\mathbf {S_{1}}\times \mathbf {{S_{3}}})\nonumber\\
&&+e_{1,4} (\mathbf {S_{1}}\times \mathbf {{S_{4}}})
+e_{1,5} (\mathbf {S_{1}}\times \mathbf {{S_{5}}}) ]\label{interf}
\end{eqnarray}

%Fig2

Since we suppose $\mathbf P_{k}$ is a vector of magnitude 1 pointing along the $z$ axis, namely its $z$ component is $P_{k}^z =\pm 1$, we will use hereafter $P_{k}^z$ for electric polarization instead of $\mathbf P_{k}$.

From Eq. (\ref{Hmf}), we see that the magnetoelectric interaction $J^{mf}$ favors a canted spin
structure. It competes with the exchange interaction $J$ of
$H_m$ which favors collinear spin configurations.  Usually the magnetic or
ferroelectric exchange interaction is the leading term in the
Hamiltonian, so that in many situations the magnetoelectric
effect is negligible. However, in nanofilms of superlattices the
magnetoelectric interaction is crucial for the creation of
non-collinear long-range spin order.

\subsection{Ground state}

\subsection{Ground state in zero magnetic field}
Let us analyze the structure of the ground state (GS) in zero magnetic field.  Since the polarizations are along the $z$ axis, the interface DM interaction is minimum when $\mathbf {S_{i}}$ and $\mathbf {S_{j}}$ lie in the $xy$ interface plane and perpendicular to each other.  However the ferromagnetic exchange interaction among the spins will compte with the DM perpendicular configuration. The resulting configuration is non collinear. We will determine it below, but at this stage, we note that the ferroelectric film has always polarizations along the $z$ axis even when interface interaction is turned on.

Let us determine the GS spin configurations in magnetic layers in zero field.  If the magnetic film has only one monolayer, the minimization of $H^{mf}$ in zero magnetic field is done as follows.

By symmetry, each spin has the same angle $\theta$ with its four NN in the $xy$ plane. The energy of the spin $\mathbf S_i$ gives the relation between $\theta$ and $J^{m}$
\begin{equation}\label{monolayer}
E_{i}=-4J^{m}S^2\cos\theta + 8J^{mf}P^zS^2\sin\theta
\end{equation}
where $\theta=|\theta_{i,j}|$ and care has been taken on the signs of $\sin \theta_{i,j}$ when counting NN, namely
two opposite NN have opposite signs, and the oppossite coefficient $e_{ij}$, as given in Eq. (\ref{interf}).  Note that the coefficient 4 of the first term is the number of in-plane NN pairs , and the coefficient 8 of the second term is due to the fact that each spin has 4 coupling DM pairs with the NN polarization in the upper ferroelectric plane, and 4 with the NN polarization of the lower ferroelectric plane (we are in the case of a magnetic monolayer). The minimization
of $E_i$ yields, taking $P^z=1$ in the GS and $S=1$,
\begin{equation}
\frac{dE_{i}}{d\theta}=0\  \ \Rightarrow \  \  \frac{-2J^{mf}}{J^m}=\tan \theta
\  \ \Rightarrow \  \  \theta=\arctan (-\frac{2J^{mf}}{J^m})\label{gsangle}
\end{equation}
The value of $\theta$ for a given $\frac{-2J^{mf}}{J^m}$ is precisely what obtained by
the numerical minimization of the energy.  We see that when $J^{mf}\rightarrow 0$, one has $\theta \rightarrow 0$, and when $J^{mf}\rightarrow -\infty$, one has $J^{mf}\rightarrow \pi/2$ as it should be.  Note that we will consider in this paper $J^{mf}<0$ so as to have $\theta >0$.

The above relation between the angle and $J^{mf}$ will be used in the next section to calculate the spin waves in the case of a magnetic monolayer sandwiched between ferroelectric films.

In the case when the magnetic film has a thickness, the angle between NN spins in each magnetic layer is different from that of the neighboring layer. It is more convenient using the numerical minimization method called "steepest descent method" to obtain the GS spin configuration.  This method consists in minimizing the energy of each spin by aligning it parallel to the local field acting on it from its NN. This is done as follows. We generate a random initial spin configuration, then we take one spin and calculate the interaction field from its NN.  We align it in the direction of this field, and take another spin and repeat the procedure until all spins are considered. We go again for another sweep until the total energy converges to a minimum. In principle, with this iteration procedure the system can be stuck in a meta-stable state when there is a strong interaction disorder
such as in spin-glasses. But for uniform, translational interactions, we have never encountered such a problem in many systems studied so far.

We use a sample size $N\times N\times L$. For most calculations, we select  $N=40$ and $L=8$ using the periodic boundary conditions in the $xy$ plane. For simplicity, when we
investigate the effect of the exchange couplings on the magnetic and
ferroelectric properties, we take the same thickness for the upper
and lower layers $L_{a}= L_{b}=4=L/2$. Exchange parameters between spins
and polarizations  are taken as $J^{m}=J^{f}=1$ for the simulation.
For simplicity we will consider the case where the in-plane and
inter-plane exchange magnetic and ferroelectric interactions between
nearest neighbors are both positive. All the results are obtained with
$J^{m}=J^{f}=1$ for different values of the interaction parameter
$J^{mf}$.

\begin{figure}[h]
\vspace{-10pt}
\begin{center}
\includegraphics[scale=0.30]{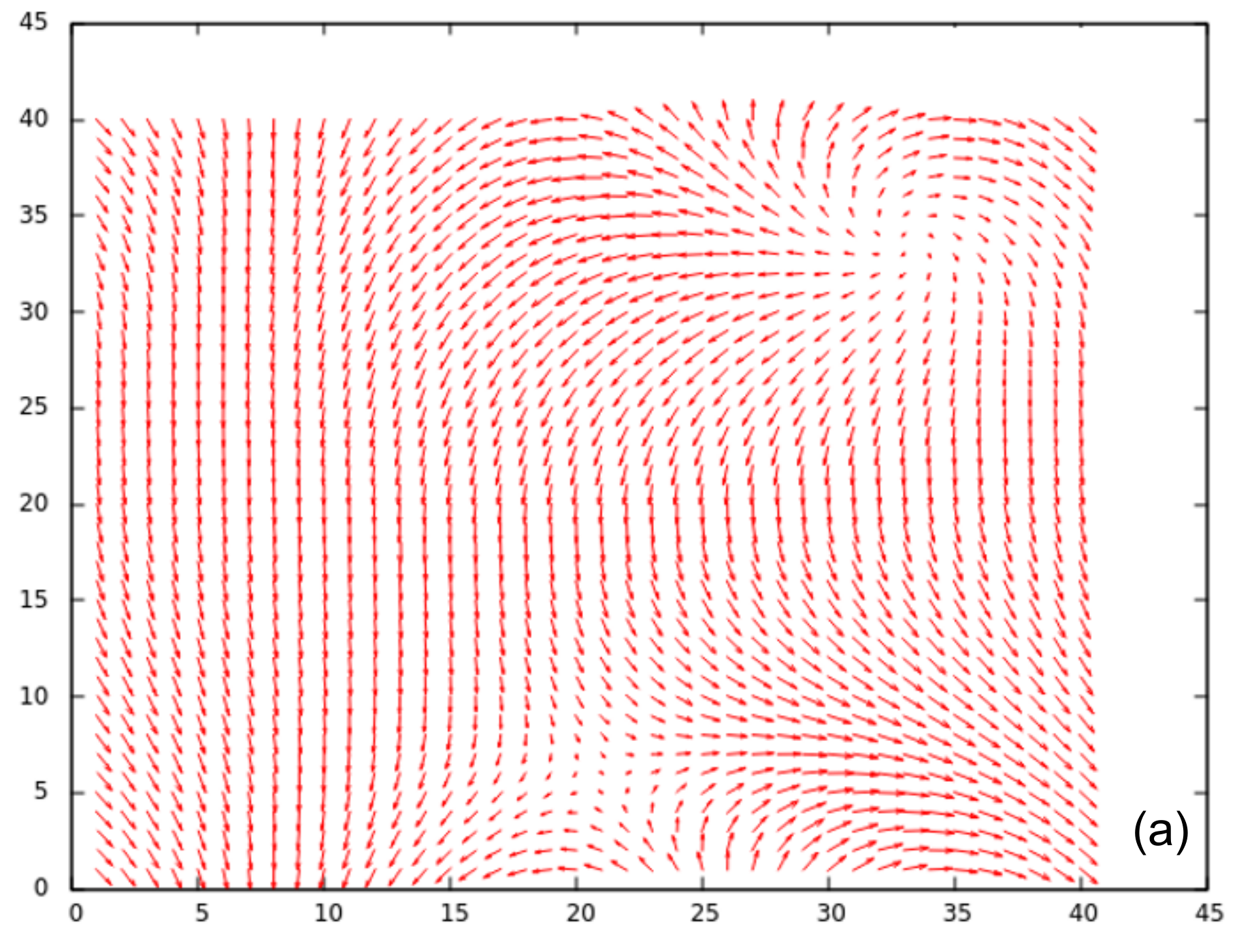}
\includegraphics[scale=0.25]{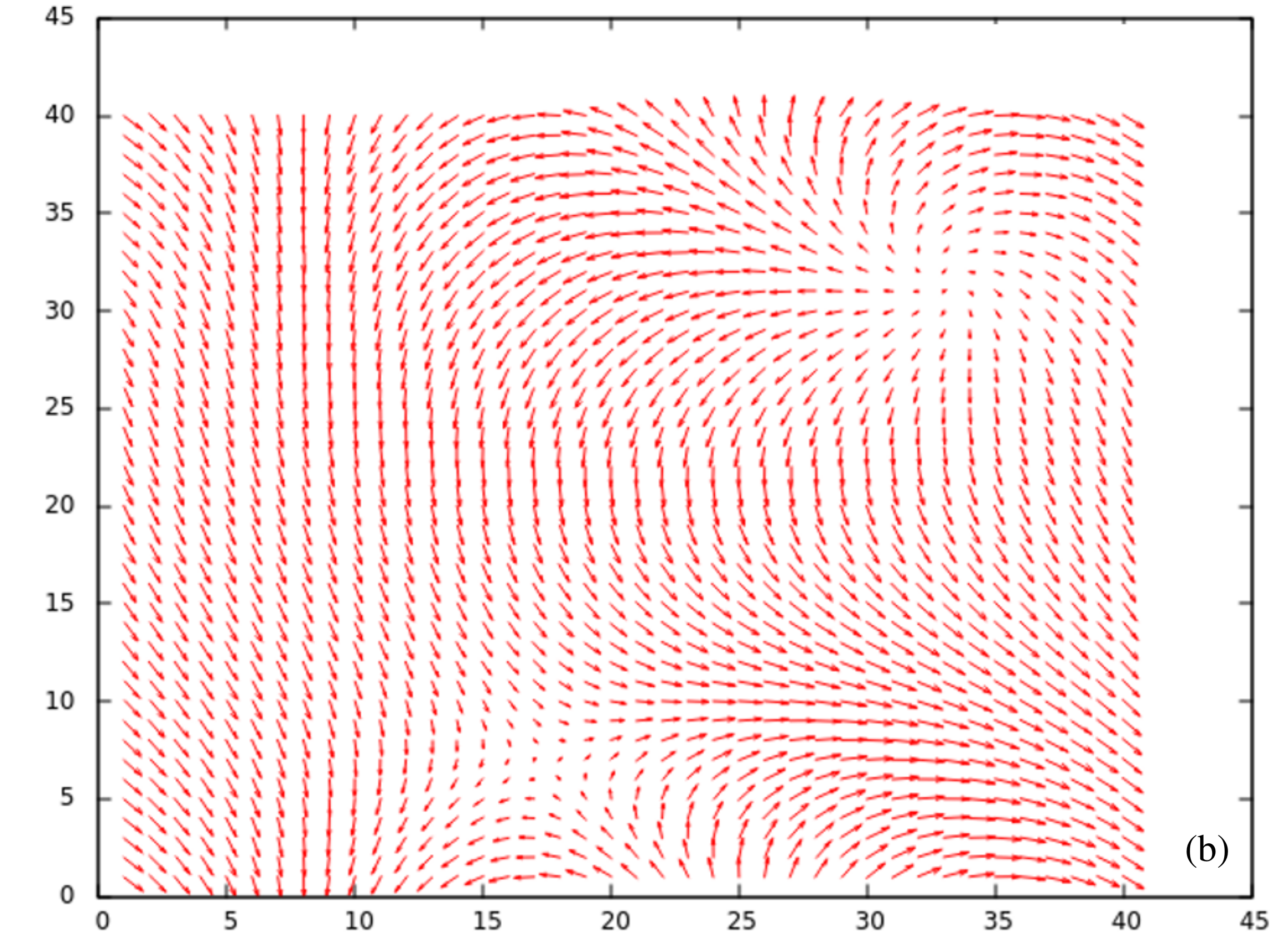}
\includegraphics[scale=0.35]{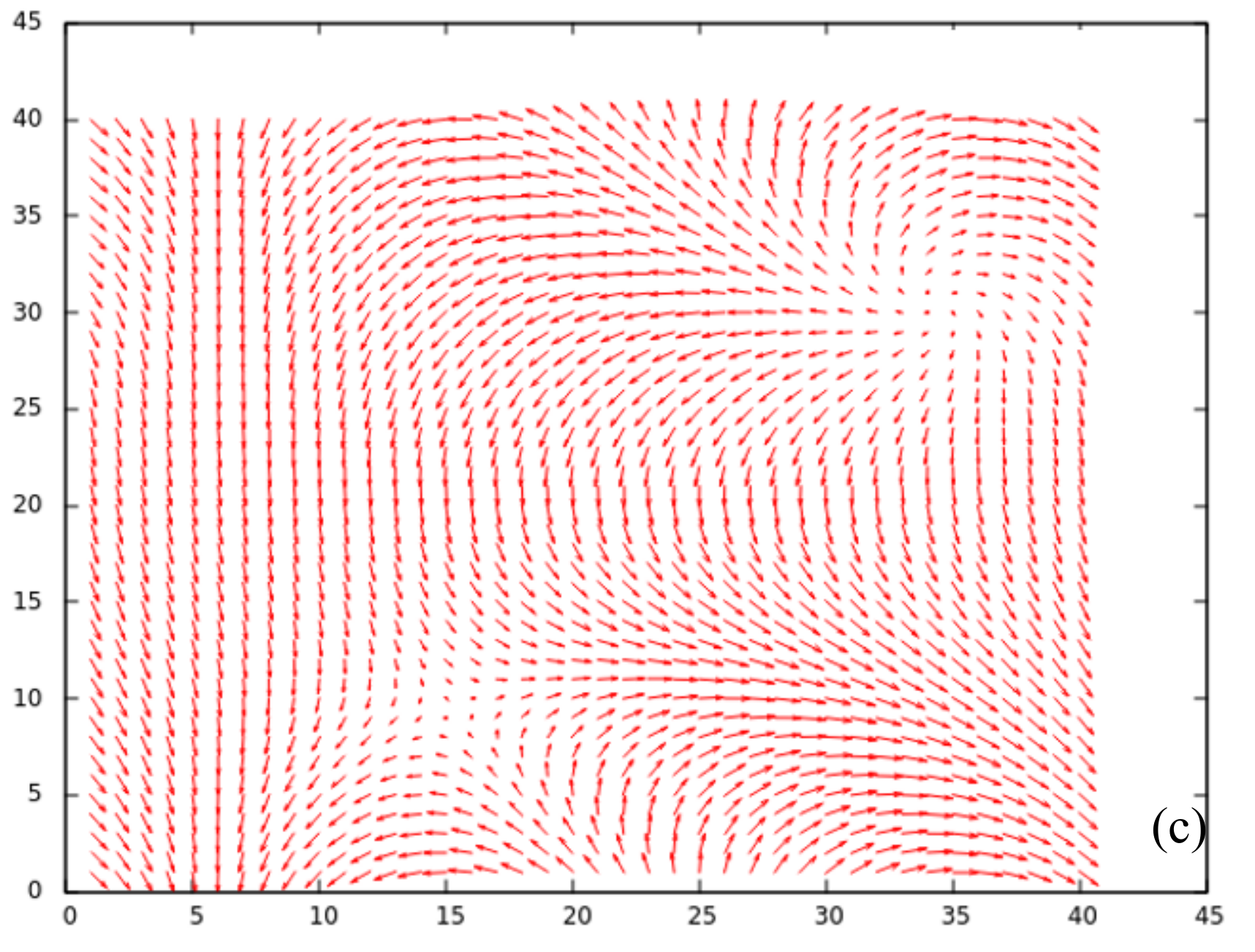}
\end{center}
\vspace{10pt} \caption{GS spin configuration for weak couplings: $J^{mf}=-0.1$ (a), -0.125 (b), -0.15 (c), with $H=0$} \label{fig4} \vspace{10pt}
\end{figure}

We investigated the following range of values for the
interaction parameters $J^{mf}$: from $J^{mf}=-0.05$ to $J^{mf}=-6.0$
with different values of the external magnetic and electric fields.
We note that the steepest descent method calculates the real ground
state with the minimum energy to the value $J^{mf}=-1.25$. After
larger values, the angle $\theta$ tends to $\pi/2$ so that all magnetic exchange terms (scalar products) will be close to zero, the minimum energy corresponds to the DM energy. Figure \ref{fig4} shows the GS configurations of the magnetic interface layer for small values of $J^{mf}$: -0.1, -0.125, -0.15. Such small values yields small values of angles between spins so that the GS configurations have ferromagnetic and non collinear domains.
Note that angles in magnetic interior layers are different but the GS configurations are of the same texture (not shown).

%Fig3

\begin{figure}[h!]
\vspace{-10pt}
\begin{center}
\includegraphics[scale=0.40]{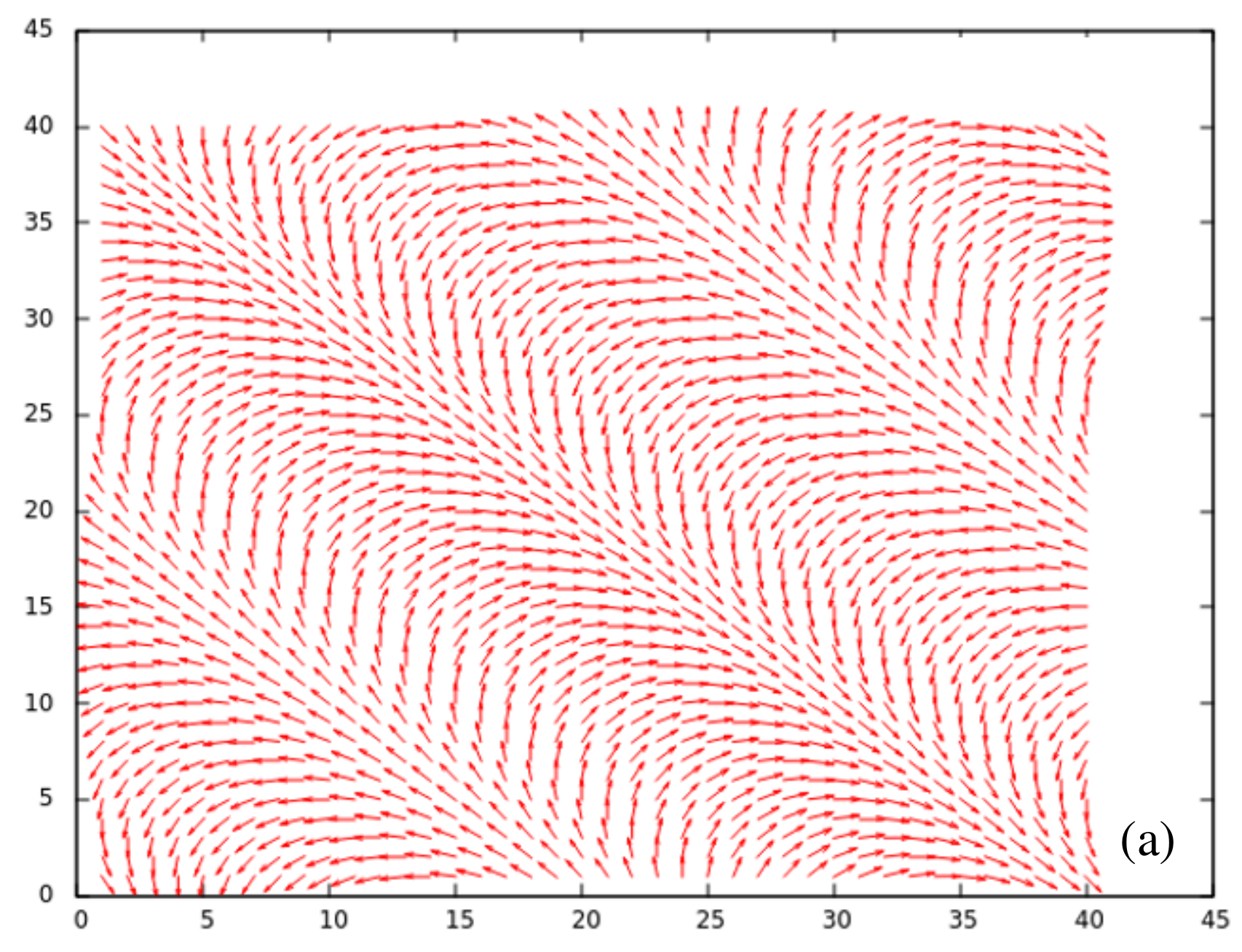}
\includegraphics[scale=0.35]{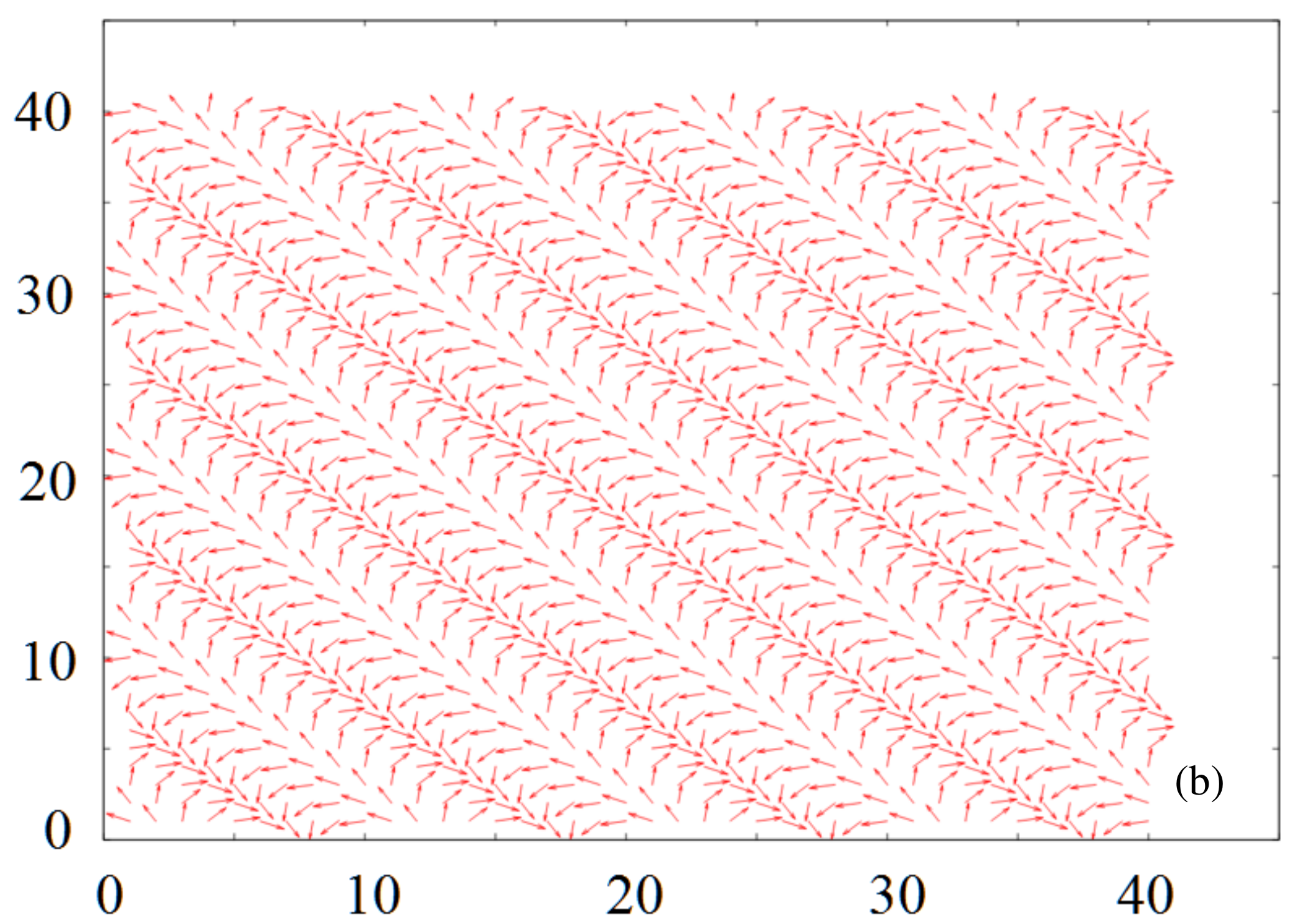}
\includegraphics[scale=0.50]{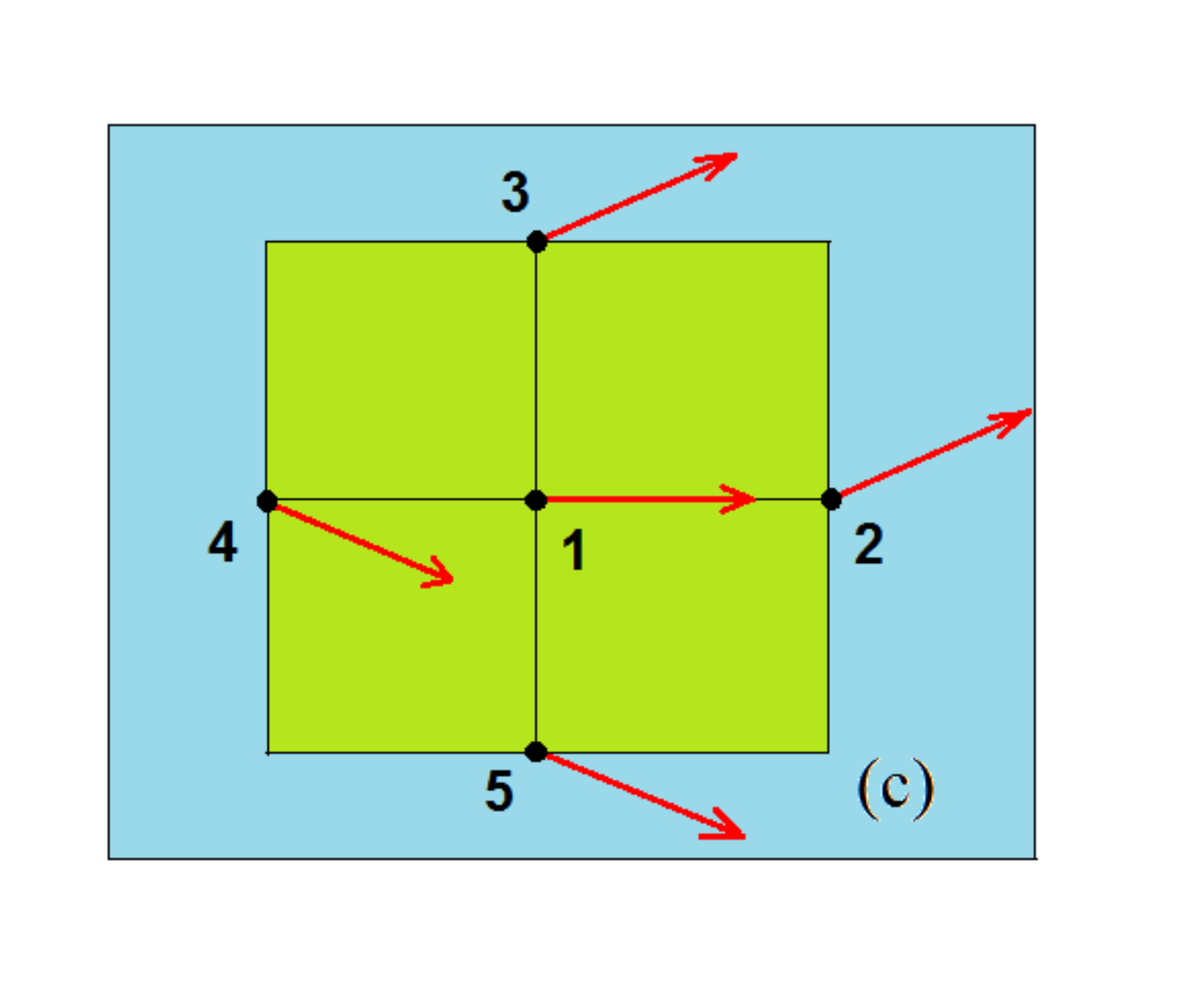}
\end{center}
\vspace{10pt} \caption{GS spin configurations for $J^{mf}=-0.45$ (a), -1.2 (b), with $H=0$. Angles between NN are schematically zoomed (c). See text for comments. } \label{fig5} \vspace{10pt}
\end{figure}

For larger values of $J^{mf}$, the GS spin configurations have periodic structures
with no more mixed domains. We show in Fig. \ref{fig5} examples where $J^{mf}=-0.45$ and -1.2. Several remarks are in order:

i) Each spin has the same turning angle $\theta$ with its NN in both $x$ and $y$ direction. The schematic zoom in Fig. \ref{fig5}c shows that the spins on the same diagonal (spins 1 and 2, spins 3 and 4) are parallel. This explains the structures shown in Figs. \ref{fig5}a and \ref{fig5}b;

ii) The periodicity of the diagonal parallel lines depends on the value of $\theta$ (comparing Fig. \ref{fig5}a and Fig. \ref{fig5}b).  With a large size  of $N$, the periodic conditions have no significant effects.

%Fig4

\subsection{Ground state in applied magnetic field}

We apply a magnetic field perpendicular to the $xy$ plane. As we know, in systems where some spin orientations are incompatible with the field such as in antiferromagnets, the down spins cannot be turned into the field direction without loosing its interaction energy with the up spins. To preserve this interaction, the spins turn into the direction almost perpendicular to the field while staying almost parallel with each other. This phenomenon is called "spin  flop" \cite{DiepTM}.  In more complicated systems such as helimagnets in a field, more complicated reaction of spins to the field was observed, leading to striking phenomena such as partial phase transition in thin helimagnetic films \cite{SahbiHeliField}.  In the present system, the

\begin{figure}[h]
\vspace{10pt}
\begin{center}
\includegraphics[scale=0.30]{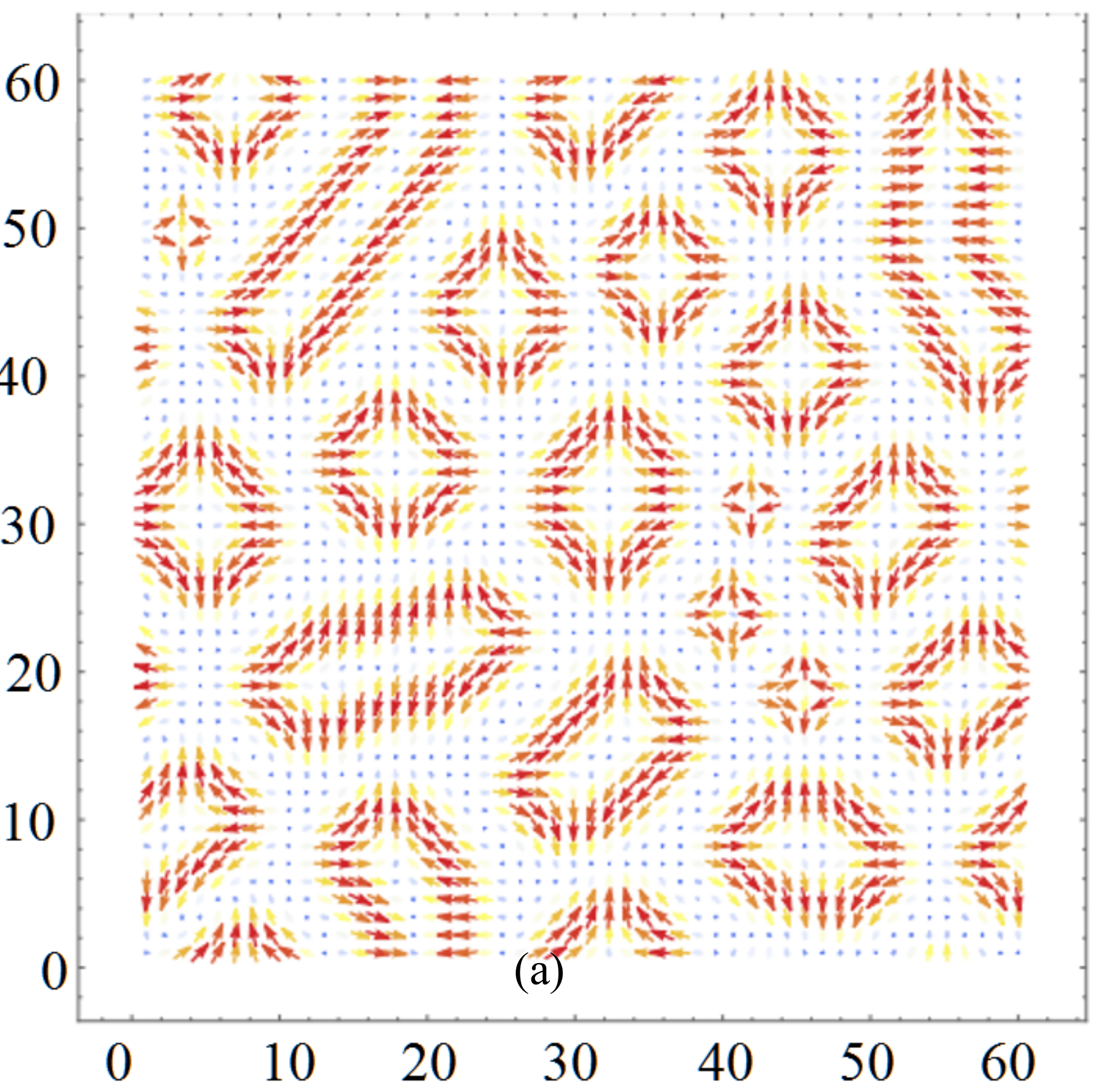}
\includegraphics[scale=0.30]{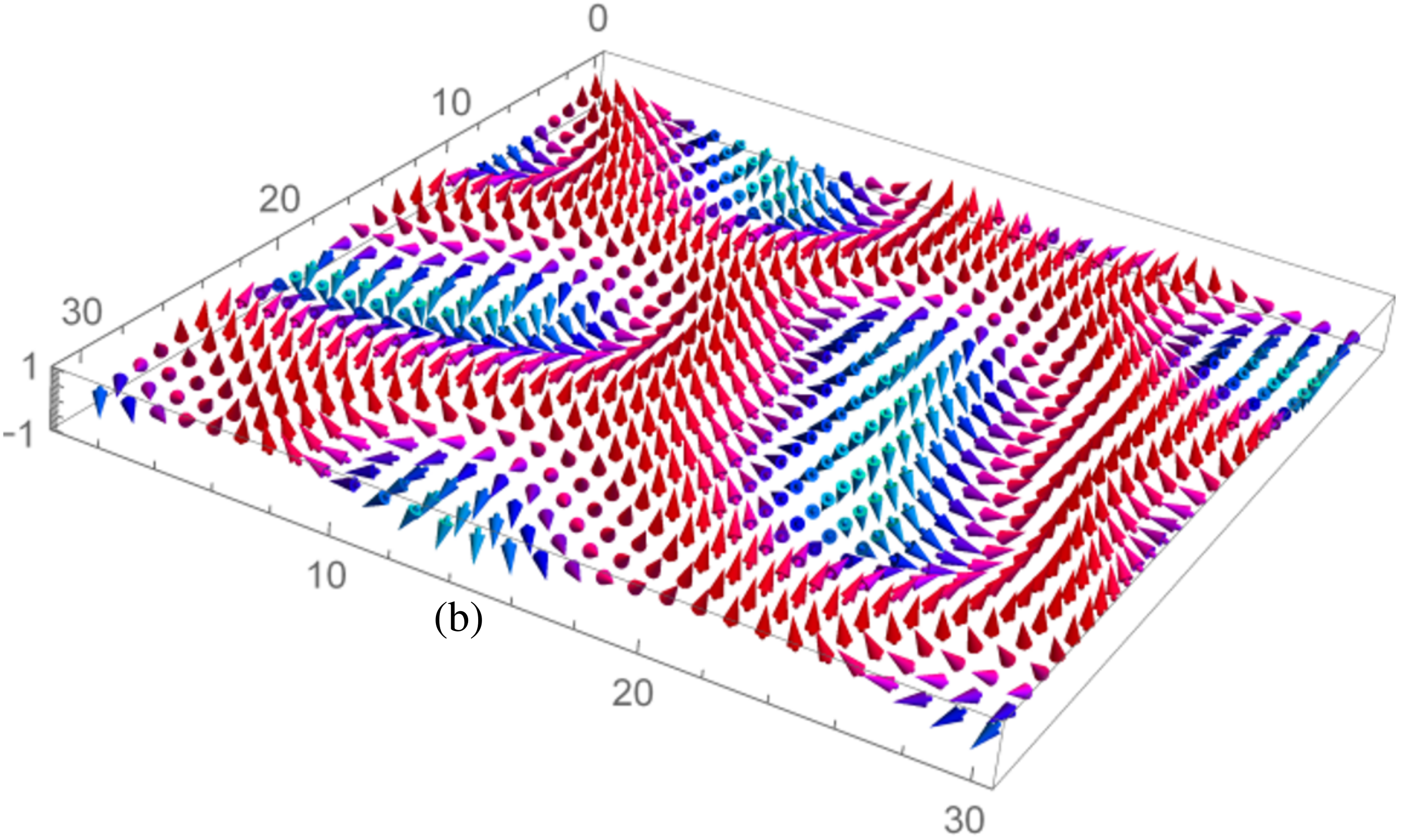}
\end{center}
\vspace{10pt} \caption{GS configuration of the surface
magnetic layer for (a) $J^{mf}=-1.1$ and $H=0.1$, (b) 3D view of the surface GS configuration.} \label{fig14}
\vspace{10pt}
\end{figure}

Figure \ref{fig14}a shows the ground state configuration for $J^{mf}=-1.1$ for
first (surface) magnetic layer, with external magnetic layer
$H=0.1$. Figure \ref{fig14}b shows the 3D view. We can observe the
beginning of the birth of skyrmions at the interface and in the interior
magnetic layer.
%Fig5

\begin{figure}[h]
\vspace{-10pt}
\begin{center}
\includegraphics[scale=0.30]{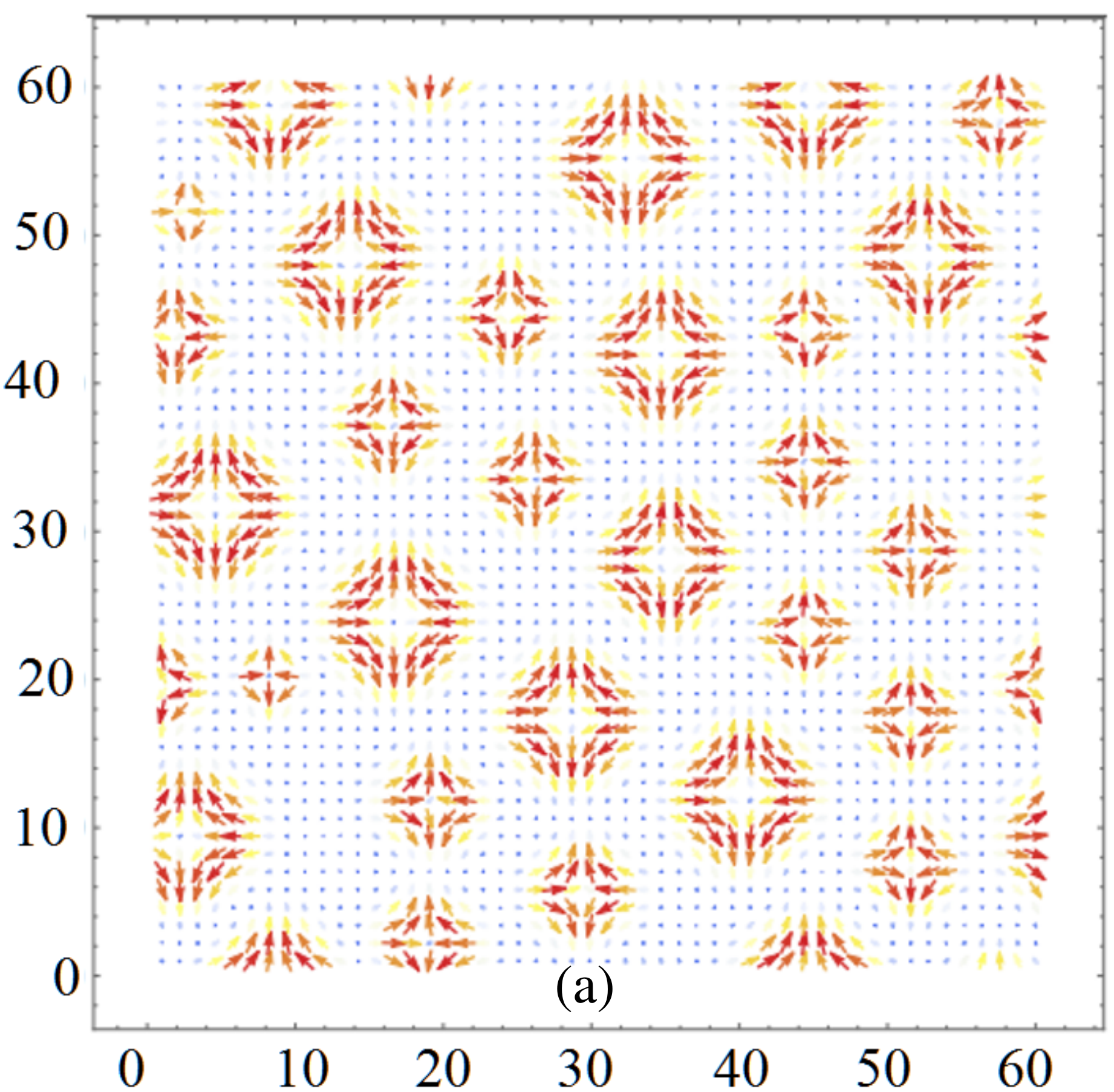}
\includegraphics[scale=0.30]{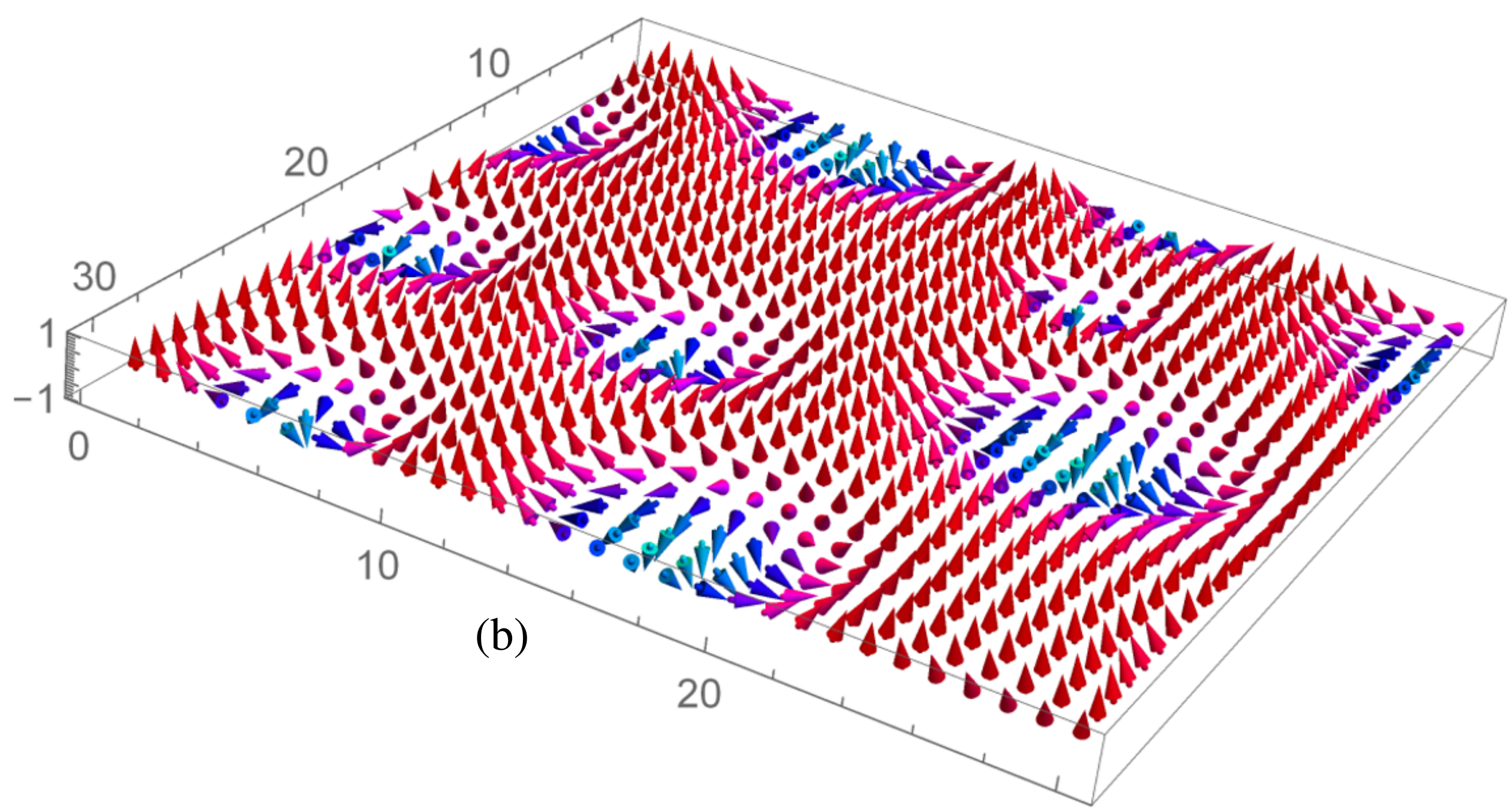}
\end{center}
\vspace{10pt} \caption{(a) GS configuration
for the surface magnetic layer for $J^{mf}=-1.1$ and $H=0.2$, (b) 3D view. } \label{fig16} \vspace{10pt}
\end{figure}

Figure \ref{fig16}a shows the ground state configuration for $J^{mf}=-1.1$ for
first (surface) magnetic layer, with external magnetic layer
$H=0.2$. Figure \ref{fig16}b shows the 3D view. We can observe the
skyrmions for the surface and interior magnetic layer.

%Fig6

\begin{figure}[h]
\vspace{-10pt}
\begin{center}
\includegraphics[scale=0.08]{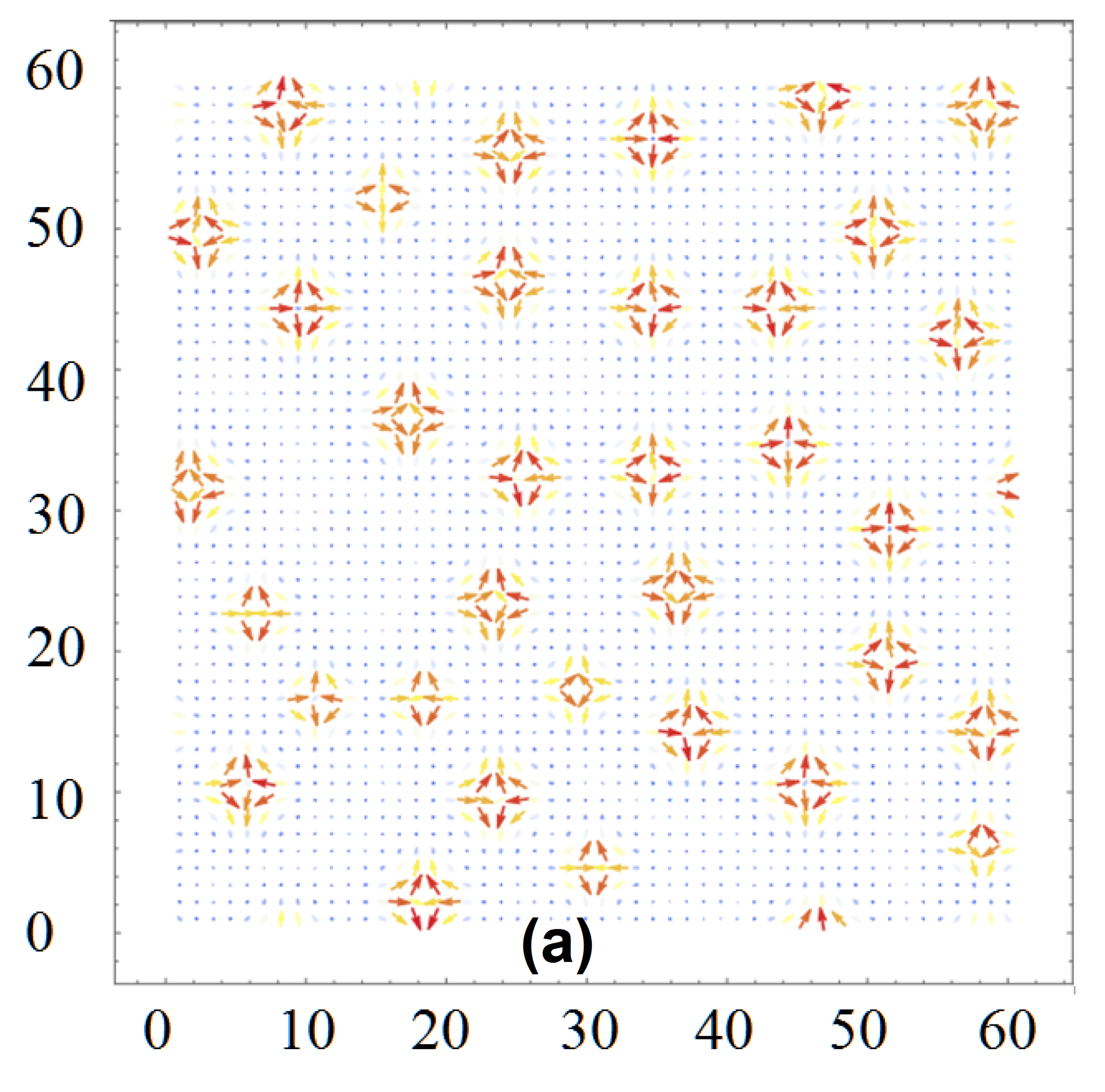}
\includegraphics[scale=0.08]{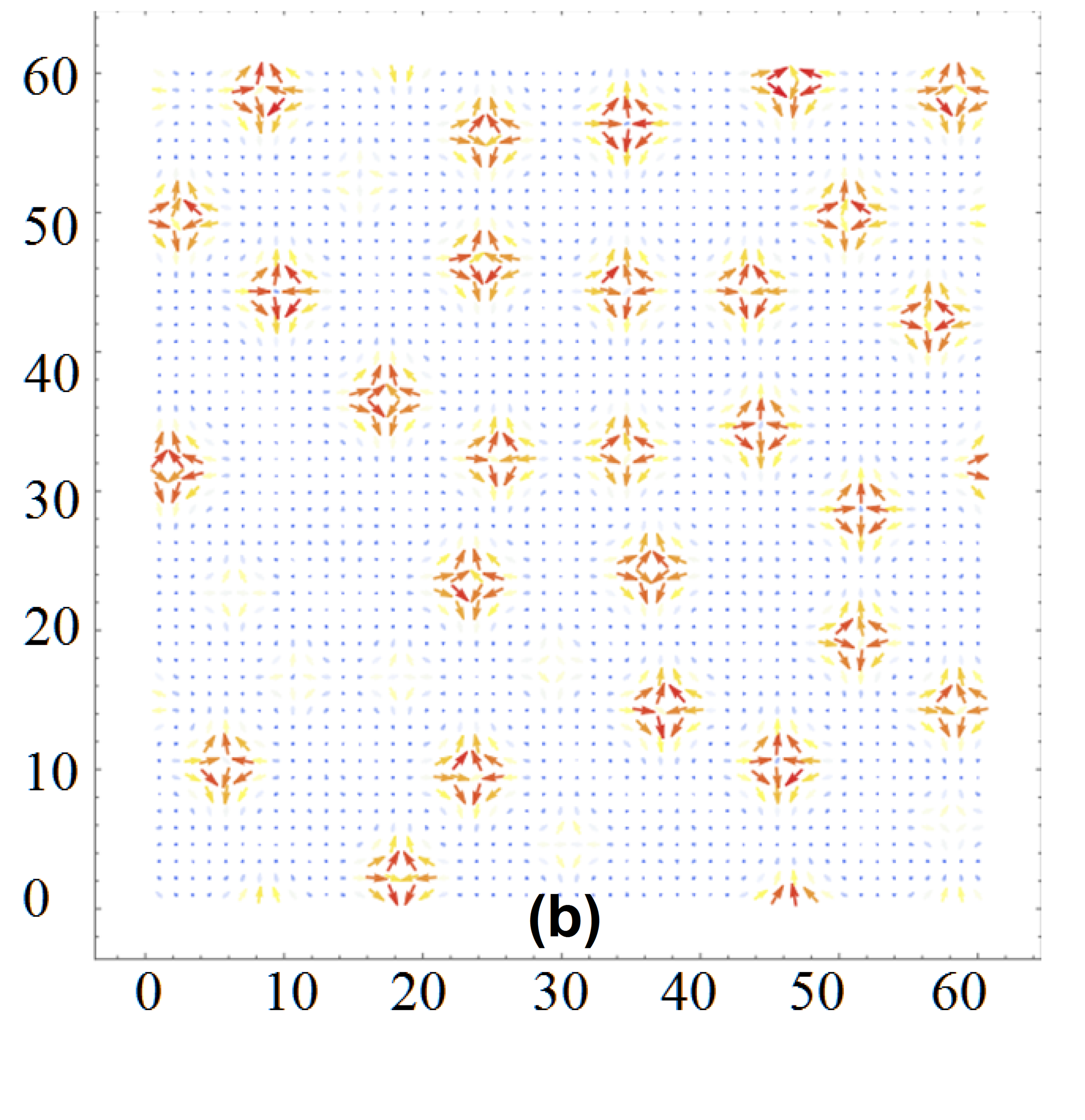}
\end{center}
\vspace{10pt} \caption{(a) GS configuration for the interface magnetic layer for $J^{mf}=-1.1$ and $H=0.33$, (b) GS configurations for the second and third magnetic layers (they are identical).  See text for comments. } \label{fig18} \vspace{10pt}
\end{figure}

Figure \ref{fig18}  shows the GS configuration of
the interface magnetic layer (top) for $J^{mf}=-1.1$, with external magnetic layer
$H=0.33$. The bottom figure shows the configurations of the second (interior) magnetic layer. We can observe skyrmions on both
the interface and the interior magnetic layers.

%Fig7

\begin{figure}[h]
\vspace{-10pt}
\begin{center}
\includegraphics[scale=0.40]{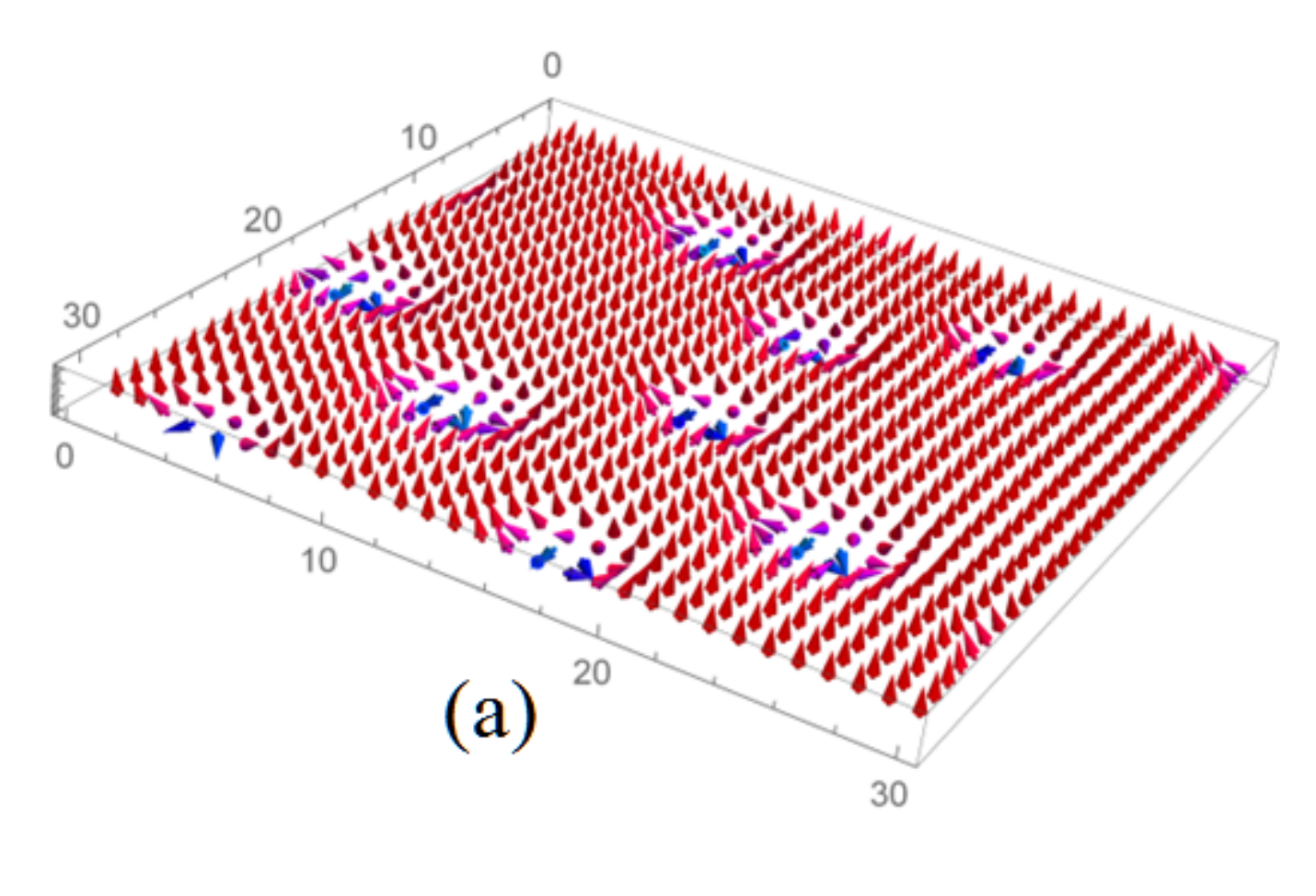}
\includegraphics[scale=0.40]{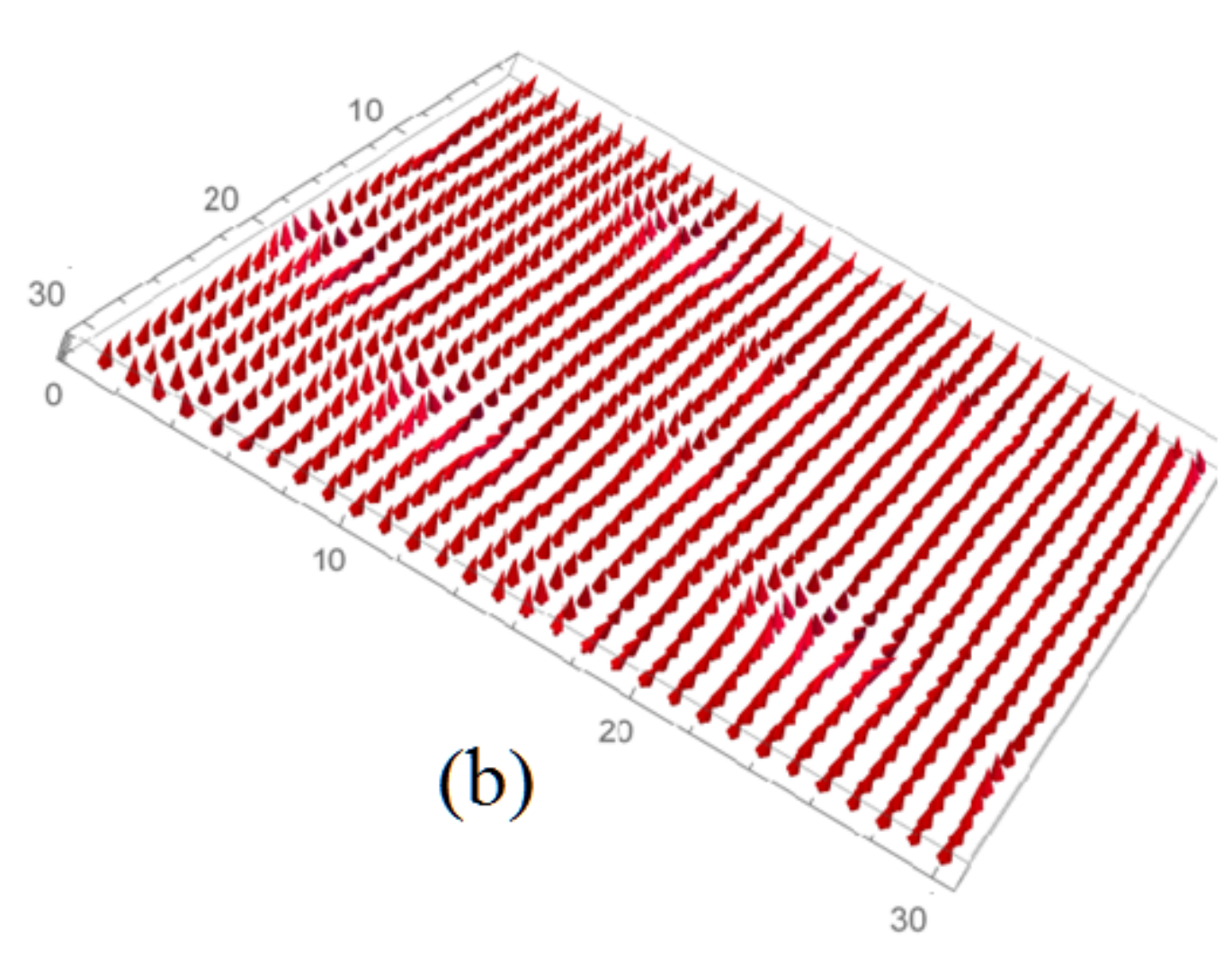}
\end{center}
\vspace{10pt} \caption{ (a) 3D view of the GS configuration of the interface, (b) 3D view of the GS configuration of the second and third magnetic layers, for $J^{mf}$ and $H=0.33$. } \label{fig20} \vspace{10pt}
\end{figure}

Figure \ref{fig20} shows the 3D view of the GS configuration for
$J^{mf}=-1.1$, with $H=0.33$  for the first (interface) magnetic layer and the second (interior) magnetic layer. We can observe skyrmions very pronounced
for the surface layer but less contrast for the interior magnetic layer. For fields stronger than $H=0.33$, skyrmions disappear in interior layers. At strong fields, all spins are parallel to the field, thus no skyrmions anywhere.

%Fig8

\section{Spin waves in zero field}\label{Green}

Before showing Monte Carlo results for the phase transition in our superlattice model, let us show theoretically  spin-waves (SW) excited in the magnetic film in zero field, in some simple cases.  The method we employ is the Green's function technique for non  collinear spin configurations which has been shown to be efficient for studying low-$T$ properties of quantum spin systems such as helimagnets \cite{PhysRevB.91.014436} and systems with a DM interaction \cite{SahbiSW}.

In this section, we consider the same Hamiltonian supposed in Eqs. (\ref{eq-ham-sysm-1})-(\ref{Hmf}) but with quantum spins of amplitude 1/2.

\begin{figure}[h]
\vspace{10pt}
\begin{center}
\includegraphics[scale=0.40]{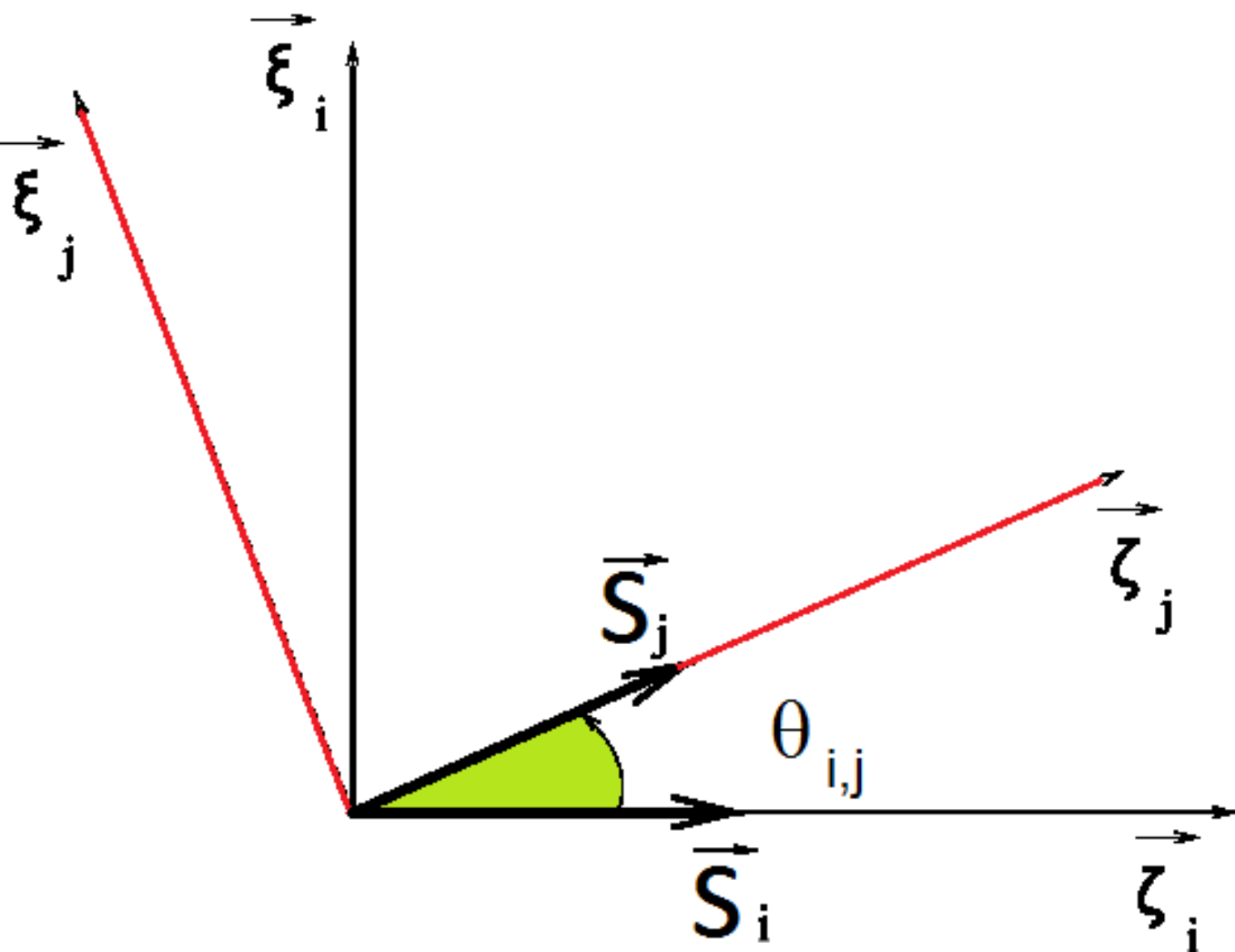}
\end{center}
\caption{The spin quantization axes of $\mathbf S_{i}$ and $\mathbf S_{j}$ are  $\hat\zeta_{i}$ and $\hat\zeta_{j}$, respectively, in the $xy$ plane. \label{ffig2}}
\end{figure}

As seen in the previous section, the spins lie in the $xy$ planes, each on its quantization
local axis lying in the $xy$ plane (quantization axis being the $\zeta$ axis, see Fig. \ref{ffig2}).

%Fig9
Expressing the spins in the local coordinates, one has
\begin{eqnarray}
\mathbf S_i&=&S_i^{\xi_i}\hat \xi_i+S_i^{\eta_i}\hat \eta_i+S_i^{\zeta_i}\hat \zeta_i\label{SI}\\
\mathbf S_j&=&S_j^{\xi_j}\hat \xi_j+S_j^{\eta_j}\hat \eta_j+S_j^{\zeta_j}\hat \zeta_j\label{SJ}
\end{eqnarray}
where the $i$ and $j$ coordinates are connected by the rotation
\begin{eqnarray}
\hat \xi_j&=&\cos \theta_{ij}\hat \zeta_i+\sin \theta_{ij}\hat \xi_i\nonumber\\
\hat \zeta_j&=&-\sin \theta_{ij}\hat \zeta_i+\cos \theta_{ij}\hat \xi_i\nonumber\\
\hat \eta_j&=&\hat \eta_i\nonumber
\end{eqnarray}
where $\theta_{ij}=\theta_i-\theta_j$ being the angle between $\mathbf S_i$ and $\mathbf S_j$.

As we have seen above, the GS spin configuration for one monolayer is periodically non collinear.  For two-layer magnetic film, the spin configurations in two layers are identical by symmetry.  However, for thickness larger than 2, the interior layer have angles different from that on the interface layer. It is not our purpose to treat that case though it is possible to do so using the method described in Ref. \onlinecite{SahbiSW}.  We rather concentrate ourselves in the case of a monolayer in this section.

In this paper, we consider the case of spin one-half $S=1/2$. Expressing the total magnetic Hamiltonian $\mathcal H_{M}=\mathcal H_m+\mathcal H_{mf}$ in the local coordinates \cite{SahbiSW}. Writing $\mathbf S_j$ in the coordinates $(\hat \xi_i,\hat \eta_i,\hat \zeta_i)$, one gets the following exchange Hamiltonian from Eqs. (\ref{eq-ham-sysm-1})-(\ref{Hmf})

\begin{eqnarray}
\mathcal H_M &=& - \sum_{<i,j>}
J^{m}\Bigg\{\frac{1}{4}\left(\cos\theta_{i,j} -1\right)
\left(S^+_iS^+_j +S^-_iS^-_j\right)\nonumber\\
&+& \frac{1}{4}\left(\cos\theta_{i,j} +1\right) \left(S^+_iS^-_j
+S^-_iS^+_j\right)\nonumber\\
&+&\frac{1}{2}\sin\theta_{i,j}\left(S^+_i +S^-_i\right)S^z_j
-\frac{1}{2}\sin\theta_{i,j}S^z_i\left(S^+_j
+S^-_j\right)\nonumber\\
&+&\cos\theta_{i,j}S^z_iS^z_j\Bigg\}\nonumber\\
&+&\frac{D}{4}\sum_{\left<i,j\right>}[(S_i^++S_i^-)(S_j^++S_j^-)|\sin\theta_{i,j}|\nonumber\\
&&+4S_i^zS_j^z|\sin\theta_{i,j}|]\nonumber\\
&&\label{eq:HGH2}
\end{eqnarray}
where $D=J^{mf}P^z$. Note that $P^z=1$ in the GS.  At finite $T$ we replace $P^z$ by $<P^z>$.
In the above equation, we have used standard notations of spin operators for easier recognition when  using the commutation relations in the course of calculation, namely

\begin{eqnarray}
(S_i^{\xi_i}, S_i^{\eta_i}, S_i^{\zeta_i})\rightarrow (S_i^{x}, S_i^{y}, S_i^{z})\nonumber\\
(S_j^{\xi_j}, S_j^{\eta_j}, S_j^{\zeta_j})\rightarrow (S_j^{x}, S_j^{y}, S_j^{z})
\end{eqnarray}
where we understand that $S_i^{x}$ is in fact $S_i^{x_i}$ and so on.

Note that the sinus terms of $\mathcal H_m$,
the 3rd line of Eq. (\ref{eq:HGH2}), are zero when summed up on opposite NN unlike the sinus term of the DM Hamiltonian $H_{mf}$, Eq. (\ref{Hmf}) which remains thanks to the choice of the DM vectors for opposite directions in Eq. \cite{SahbiSW}.

\subsection{Monolayer}

In two dimensions (2D) there is no long-range order at finite temperature ($T$) for isotropic
spin models with short-range interaction \cite{Mermin}.
Therefore to stabilize the ordering at finite $T$ it is
useful to add an anisotropic interaction.
We use the following anisotropy between $\mathbf S_i$
and $\mathbf S_j$ which stabilizes the angle determined above between their local quantization axes $S^z_i$ and $S^z_j$:
\begin{equation}
\mathcal H_a= -\sum_{<i,j>} K_{i,j}S^z_iS^z_j\cos\theta_{i,j}
\end{equation}
where $K_{i,j}$ is supposed to be positive, small compared to $J^{m}$, and limited to NN.
Hereafter we take $I_{i,j}=K$ for NN pair in the $xy$ plane,
for simplicity.   The total magnetic Hamiltonian $\mathcal H_{M}$ is finally given by (using operator notations)
\begin{equation}\label{totalH}
\mathcal H_{M}=\mathcal H_m+\mathcal H_{mf}+\mathcal H_a
\end{equation}

We now define the following two double-time Green's functions in the real space
\begin{eqnarray}
G_{i,j}(t,t')&=&<<S_i^+(t);S_{j}^-(t')>>\nonumber\\
&=&-i\theta (t-t')
<\left[S_i^+(t),S_{j}^-(t')\right]> \label{green59a}\\
F_{i,j}(t,t')&=&<<S_i^-(t);S_{j}^-(t')>>\nonumber\\
&=&-i\theta (t-t')
<\left[S_i^-(t),S_{j}^-(t')\right]>\label{green60}
\end{eqnarray}
The equations of motion of these functions
read
\begin{eqnarray}
i\hbar\frac{dG_{i,j}(t,t')}{dt}&=&<\left[S_i^+(t),S_{j}^-(t')\right]>\delta(t-t')\nonumber\\
&&-<<\left[\mathcal H_M,S_i^+\right];S_j^->>
\label{green59a}\\
i\hbar\frac{dF_{i,j}(t,t')}{dt}&=&<\left[S_i^-(t),S_{j}^-(t')\right]>\delta(t-t')\nonumber\\
&&-<<\left[\mathcal H_M,S_i^-\right];S_j^->>
\label{green60}
\end{eqnarray}
For the $\mathcal H_m$ and $\mathcal H_a$ parts, the above equations of motion generate terms such as
$<<S_l^zS_i^{\pm};S_j^->>$ and $<<S_l^{\pm}S_i^{\pm};S_j^->>$. These functions can be approximated
by using the Tyablikov decoupling to reduce to the above-defined $G$ and $F$ functions:
\begin{eqnarray}
&&<<S_l^zS_i^{\pm};S_j^->>\simeq <S_l^z><<S_i^{\pm};S_j^->>\label{Tya1}\\
&&<<S_l^{\pm}S_i^{\pm};S_j^->>\simeq <S_l^{\pm}> <<S_i^{\pm};S_j^->>\simeq 0\label{Tya2}
\end{eqnarray}
The last expression is due to the fact that transverse spin-wave motions $<S_l^{\pm}>$ are zero with time.
For the DM term, the commutation relations $[\mathcal H,S_i^{\pm}]$ give rise to the following term:
\begin{equation}
D\sum_l\sin \theta [\mp S_i^z(S_l^{+}+S_l^{-})\pm 2S_i^{\pm} S_l^z]\label{HDMC}
\end{equation}
This leads to the following type of Green's function:
\begin{equation}
<<S_i^zS_l^{\pm};S_j^->>\simeq <S_i^z><<S_l^{\pm};S_j^->>\label{DMa}\\
\end{equation}
Note that we have used defined $\theta$ positively.
The above equation is thus related to $G$ and $F$ functions [see Eq. (\ref{Tya2})].

 We use the following Fourier transforms in the $xy$ plane  of the $G$ and $F$ Green's functions:
\begin{eqnarray}
G_{i,j}(t,t',\omega)&=&\frac{1}{\Delta} \int_{BZ} d{\mathbf k_{xy}}
\mbox{e}^{-i\hbar \omega(t-t')}
 g(\omega,\mathbf k_{xz})\mbox{e}^{i\mathbf k_{xy}.(\mathbf R_i-\mathbf R_{j})}\\
F_{i,j}(t,t',\omega)&=&\frac{1}{\Delta} \int_{BZ} d{\mathbf k_{xy}}
\mbox{e}^{-i\hbar \omega(
t-t')} f(\omega,\mathbf k_{xy})\mbox{e}^{i\mathbf k_{xy}.(\mathbf R_i-
\mathbf R_{j})}
\end{eqnarray}
where the integral is performed in the first $xy$ Brillouin zone (BZ) of surface $\Delta$ and $\omega$ is the SW frequency. Let us define the SW energy as $E=\hbar \omega$ in the following.

For a monolayer, we have after the Fourier transforms
\begin{eqnarray}
(E+A)g+Bf&=&2< S^z>\nonumber\\
-Bg+(E-A)f&=&0
\end{eqnarray}
where $A$ and $B$ are

\begin{eqnarray}
A &=& -J^{m}[8<S^z>\cos\theta (1+d)
- 4 <S^z>\gamma (\cos\theta+1)]\nonumber\\
&&-4D\sin \theta < S^z>\gamma
+8D\sin \theta < S^z>\label{anterm}\\
B &=& 4J^{m} < S^z> \gamma (\cos\theta-1)
-4D \sin \theta < S^z>\gamma\label{bnterm}
\end{eqnarray}
where the reduced anisotropy is $d=K/J^m$ and  $\gamma=(\cos k_xa+\cos k_ya)/2$, $k_{x}$ and  $k_{y}$
being the wave-vector components in the $xy$ planes, $a$ the lattice constant.

The SW energies are determined by the secular equation
\begin{eqnarray}
&&(E+A)(E-A)+B^2=0\nonumber\\
&&[E+A][E-A]+B^2=0\nonumber\\
&&E^2-A^2+B^2=0\nonumber\\
&&E=\pm \sqrt{(A+B)(A-B)}\label{SWE}
\end{eqnarray}
where $\pm$ indicate the left and right SW precessions.
We see that
\begin{itemize}

 \item if $\theta=0$, we have  $B$ and the last two terms of $A$ are zero. We recover then the ferromagnetic SW dispersion relation

\begin{equation}
E=2ZJ^{m}<S^z>(1-\gamma)
\end{equation}
where $Z=4$ is the coordination number of the
square lattice (taking $d=0$),

\item  if $\theta=\pi$, we have $A=8J^{m}<S^z>$ and $B=-8J^{m}<S^z>\gamma$.
We recover then the antiferromagnetic SW energy
\begin{equation}
E=2ZJ^{m}<S^z>\sqrt{1-\gamma^2}
\end{equation}

 \item in the presence of a DM interaction, we have $0<\cos \theta < 1$ ($0<\theta<\pi/2$). If $d=0$, the
quantity in the square root of Eq. (\ref{SWE}) is always $\geq 0$ for any $\theta$. It is zero at $\gamma=1$.
We do not need an anisotropy $d$ to stabilize the SW at $T=0$. If $d=\neq zero$ then it gives a gap at $\gamma=1$.
\end{itemize}

\begin{figure}[h!]
\vspace{10pt}
\center
\includegraphics[width=6cm]{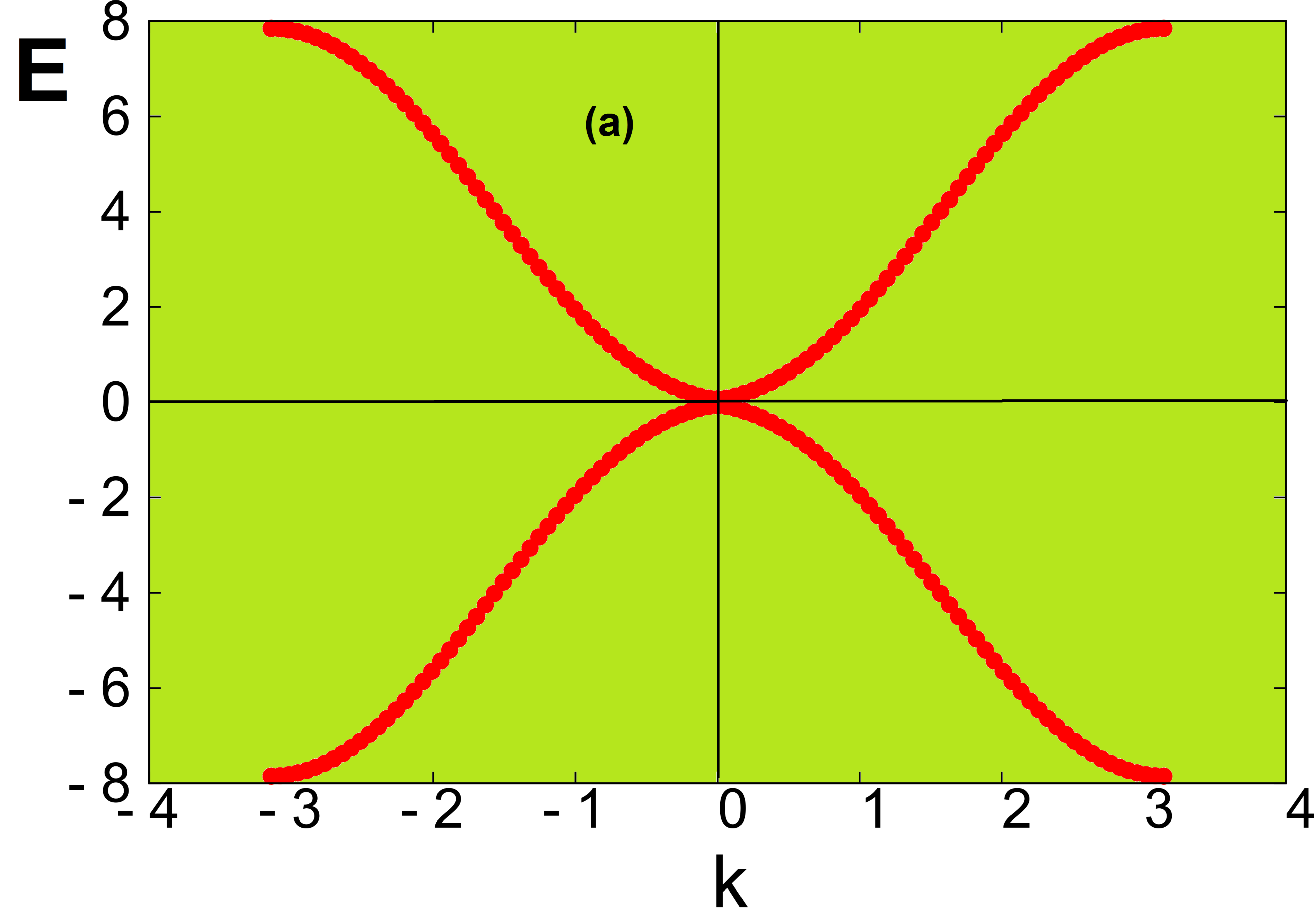}
\includegraphics[width=6cm]{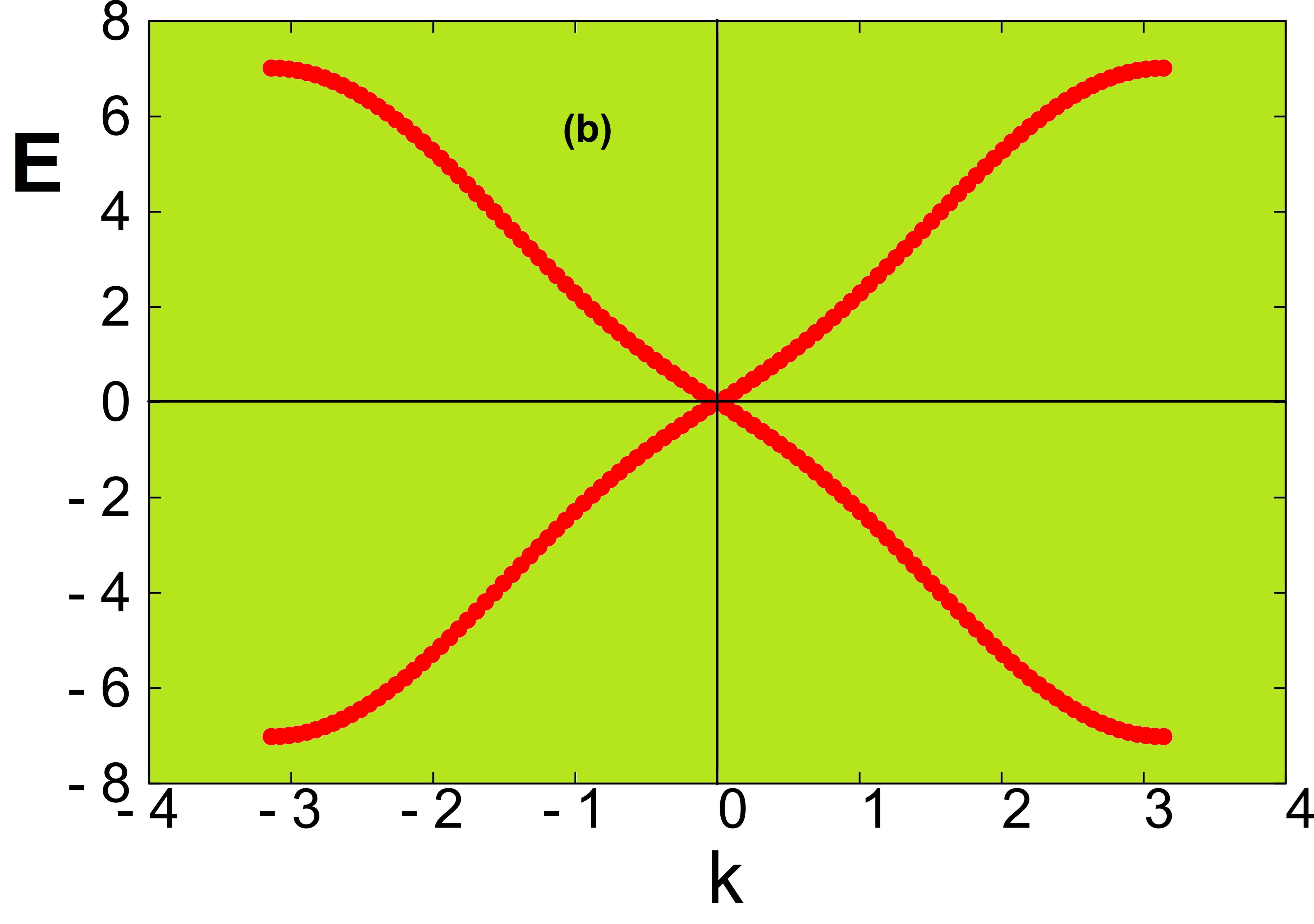}
\caption{Spin-wave energy $E (k)$ versus $k$ ($k \equiv k_x=k_z$) for (a) $\theta=0.3$ radian and (b) $\theta=1$ in 2D at $T=0$.
See text for comments.
\label{ffig3}}
\end{figure}

We show in Fig. \ref{ffig3} the SW energy calculated from Eq. (\ref{SWE}) for $\theta=0.3$ radian ($\simeq 17.2$ degrees) and 1 radian ($\simeq 57.30$ degrees). The spectrum is symmetric for positive and negative wave vectors and for left
and right precessions. Note that for small values of $\theta$ (i. e. small $D$) $E$ is proportional to $k^2$ at low $k$ (cf. Fig. \ref{ffig3}a), as in ferromagnets. However, for strong $\theta$,  $E$ is proportional to $k$ as seen in Fig. \ref{ffig3}b. This behavior is similar to that in antiferromagnets \cite{DiepTM}.  The change of behavior is progressive with increasing $\theta$, no sudden transition from $k^2$ to $k$ behavior is observed.

%Fig10

In the case of $S=1/2$, the magnetization is given by (see technical details in Ref. \onlinecite{DiepTM}):
\begin{equation}\label{lm2}
\langle S^z\rangle=\frac{1}{2}-
   \frac{1}{\Delta}
   \int
   \int dk_xdk_y
   [\frac{1}
   {\mbox{e}^{E_i/k_BT}-1}+\frac{1}
   {\mbox{e}^{-E_i/k_BT}-1}]
\end{equation}
where for each $\mathbf k$ one has $\pm E_i$ values.

Since $E_i$ depends on $S^z$, the magnetization can be calculated at finite temperatures self-consistently using the above formula.

\begin{figure}[ht!]
\centering
\includegraphics[width=7cm,angle=0]{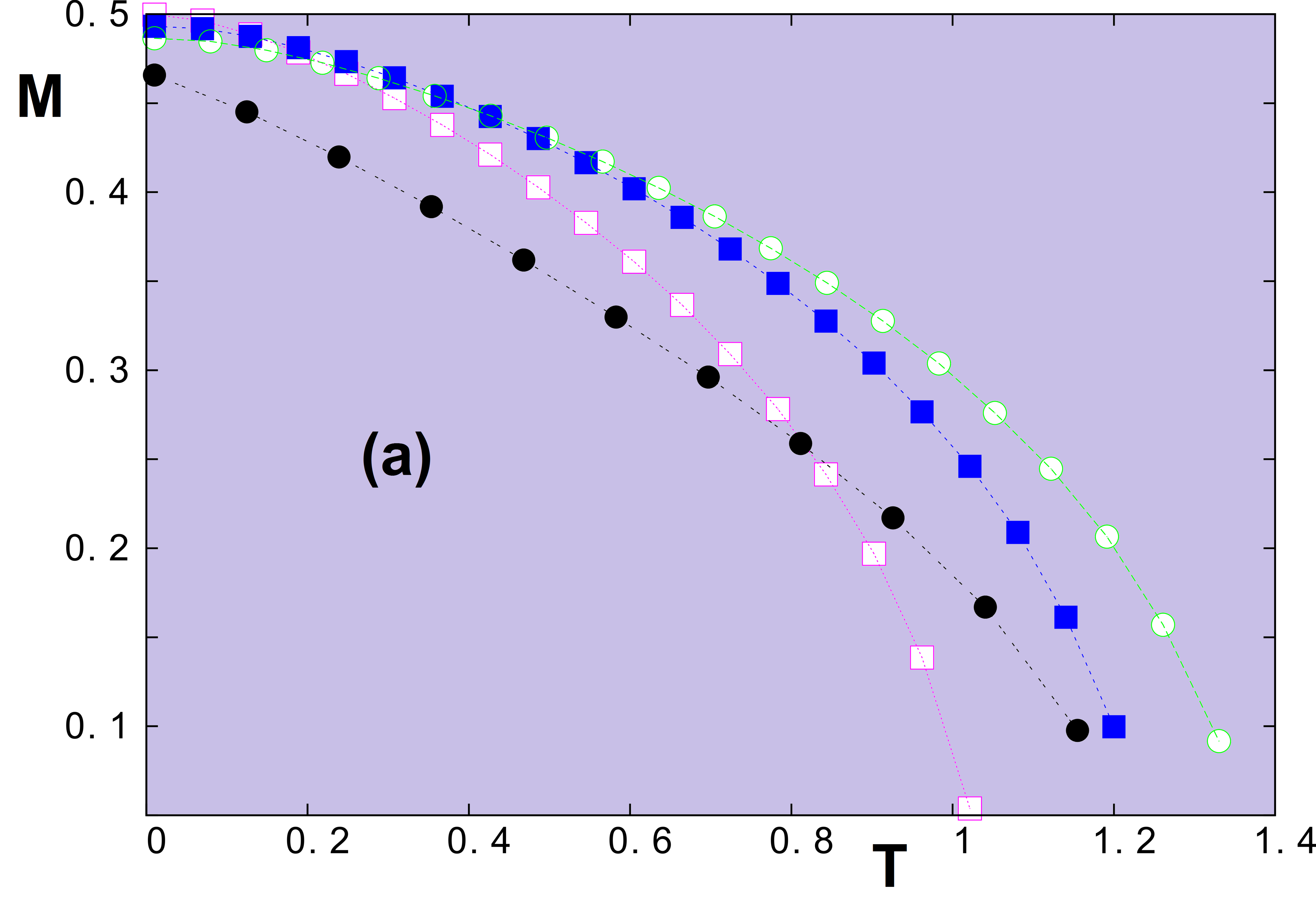}
\includegraphics[width=7cm,angle=0]{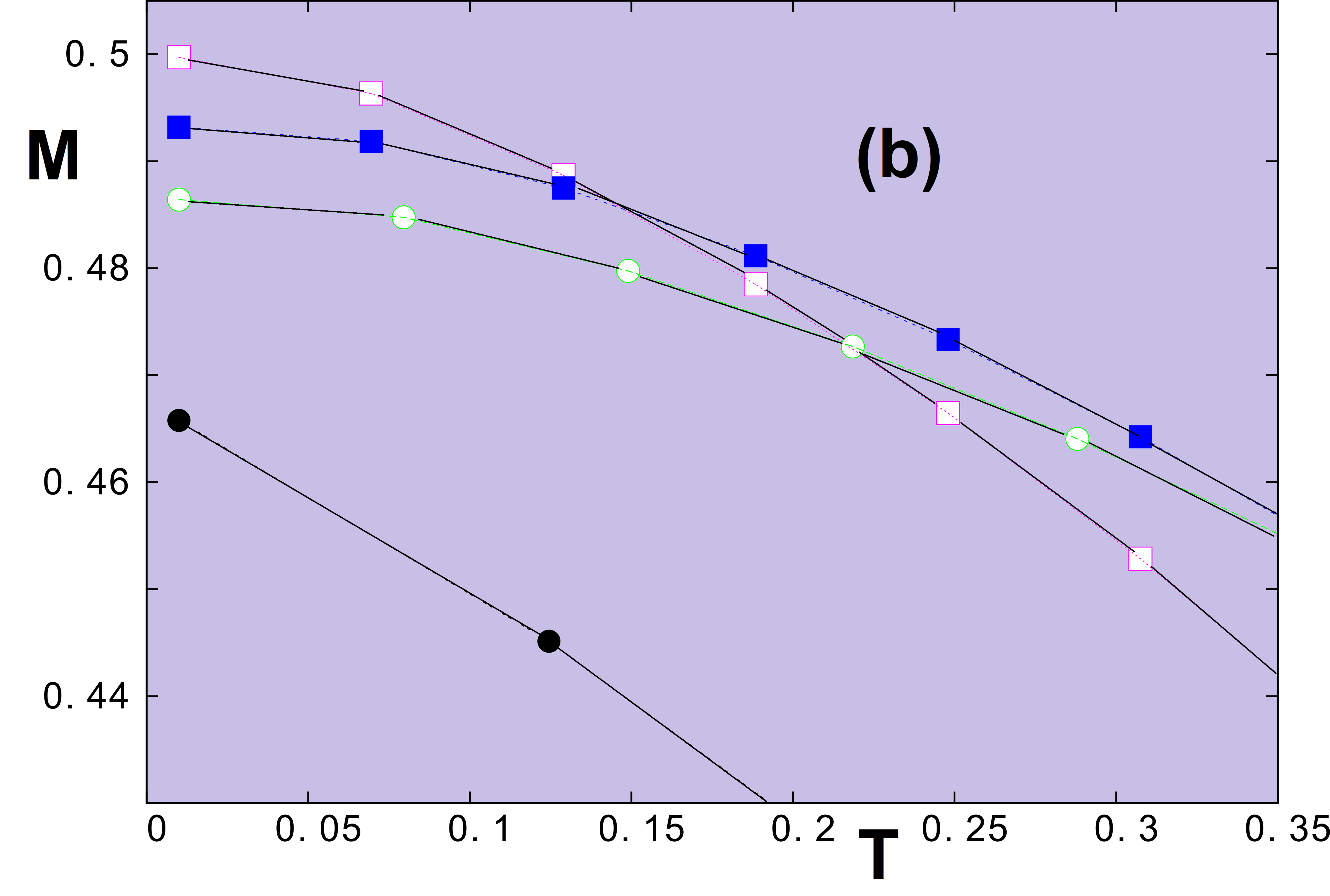}
\caption{(a) Spin length $M=<S^z>$ versus temperature $T$ for a 2D sheet with $\theta=0.175$ (radian) (magenta void  squares), $\theta=0.524$ (blue filled squares), $\theta=0.698$ (green void circles),  $\theta=1.047$ (black filled circles); (b) Zoom at low $T$ to show magnetization cross-overs.
\label{ffig4}}
\end{figure}

It is noted that the anisotropy $d$ avoids the logarithmic divergence at $k=0$ so that we can observe a long-range ordering at finite $T$
in 2D. We show in Fig. \ref{ffig4} the magnetization $M$ ($\equiv <S^z>$) calculated by Eq. (\ref{lm2}) for using $d=0.001$. It is interesting to observe that $M$ depends strongly on $\theta$: at high $T$, larger $\theta$ yields stronger $M$. However, at $T=0$ the spin length is smaller for larger $\theta$ due to the so-called spin contraction in antiferromagnets \cite{DiepTM}.  As a consequence there is a cross-over of  magnetizations with different $\theta$ at low $T$ as shown in Fig. \ref{ffig4}.

%Fig11

\begin{figure}[ht!]
\centering
\includegraphics[width=7cm,angle=0]{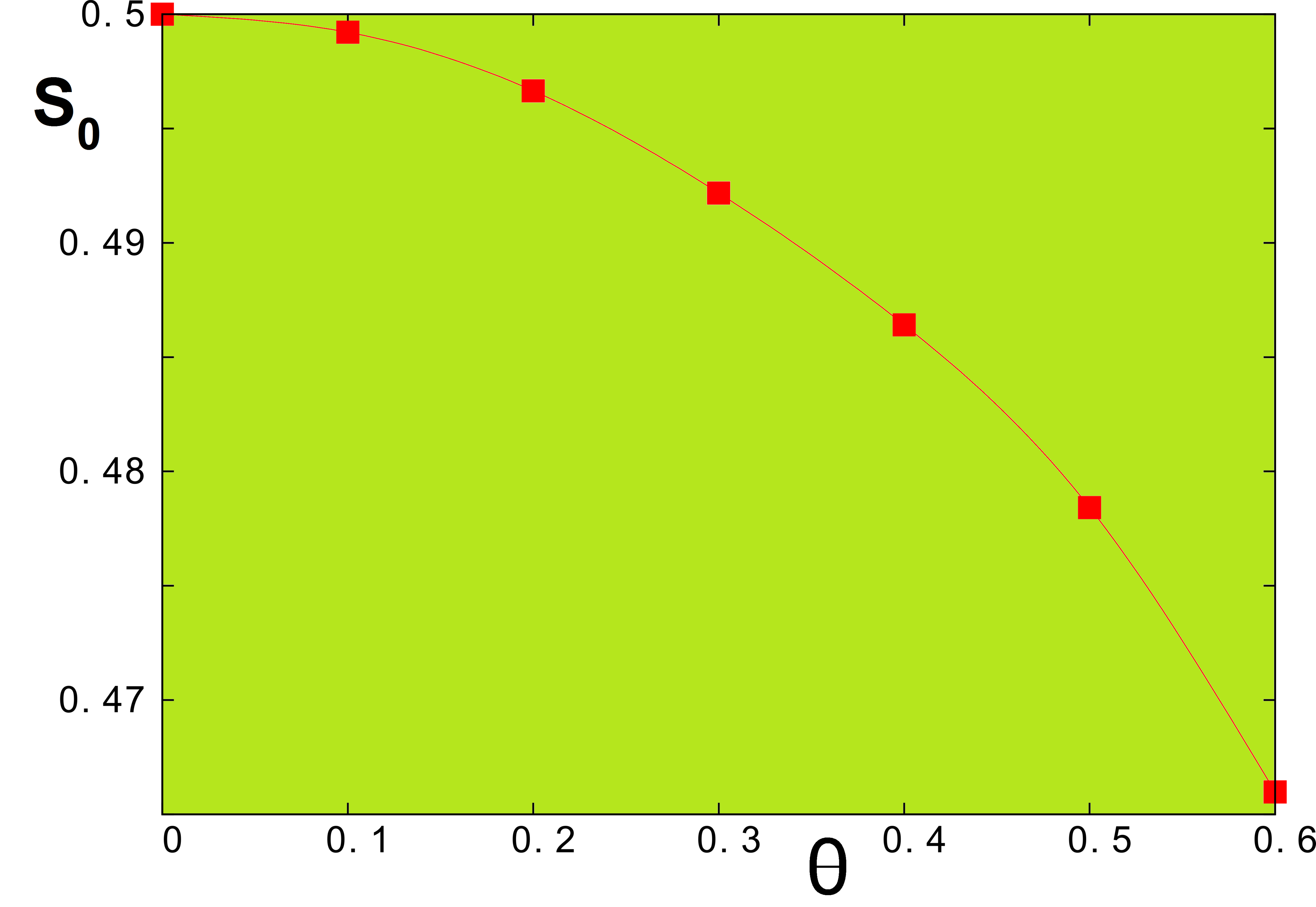}
\caption{Spin length at temperature $T=0$ for a monolayer versus $\theta$ (radian).
\label{ffig5}}
\end{figure}

The spin length at $T=0$ is shown in Fig. \ref{ffig5} for several $\theta$.

%Fig12

\subsection{Bilayer}

We note that for magnetic bilayer between two ferroelectric films, the calculation similar to that of a monolayer can be done. By symmetry, spins between the two layers are parallel, the energy of a spin on a layer is

\begin{equation}\label{bilayer}
E_{i}=-4J^{m}S^2\cos\theta -J^m S^2 + 4J^{mf}P^zS^2\sin\theta
\end{equation}
where there are 4 in-plane NN and one parallel NN spin on the other layer. The interface coupling is with only one polarization instead of two (see Eq. (\ref{monolayer})) for a monolayer for comparison.

The minimum energy corresponds to $\tan \theta=-J^{mf}/J^m$.

The calculation by the Green's functions for a film with a thickness is straightforward: writing the Green's functions for each layer and making Fourier transforms in the $xy$ planes, we obtain a system of coupled equations. For the details, the reader is referred to Ref. \onlinecite{PhysRevB.91.014436}.  For a bilayer, the SW energy is the eigenvalues of the following matrix equation
\begin{equation}
\mathbf M \left( E \right) \mathbf h = \mathbf u,
\label{eq:HGMatrix}
\end{equation}
where
\begin{equation}
\mathbf h = \left(%
\begin{array}{c}
  g_{1,n'} \\
  f_{1,n'} \\
  g_{2,n'} \\
  f_{2,n'} \\
  \end{array}%
\right) , \hspace{1cm}\mathbf u =\left(%
\begin{array}{c}
  2 \left< S^z_1\right>\delta_{1,n'}\\
  0 \\
  2 \left< S^z_{2}\right>\delta_{2,n'}\\
  0 \\
\end{array}%
\right) , \label{eq:HGMatrixgu}
\end{equation}
where $E=\hbar \omega$ and $\mathbf M\left(E\right)$ is given by
%MODIF

%\begin{widetext}
\begin{equation}
\left(%
\begin{array}{cccc}
  E+A_1& B_1    & C_1& 0\\
   -B_1   & E-A_1  & 0 & -C_1\\
  C_{2}   & 0   &E + A_{2}&B_{2}\\
  0  & -C_{2}& -B_{2}& E-A_{2}\\
\end{array}%
\right)
\end{equation}\label{eq:HGMatrixM}
%\end{widetext}
with
\begin{eqnarray}
A_{1} &=& -J^{m}[8<S^z_1>\cos\theta (1+d)
- 4 <S^z_1>\gamma (\cos\theta+1)]\nonumber\\
&&-2J^{m} <S^z_{2}>
-4D\sin \theta < S^z_{1}>\gamma
+8D\sin \theta < S^z_{1}>\label{aterm}\\
A_{2} &=& -J^{m}[8<S^z_2>\cos\theta (1+d)
- 4 <S^z_2>\gamma (\cos\theta+1)]\nonumber\\
&&-2J^{m} <S^z_{1}>
-4D\sin \theta < S^z_{2}>
+8D\sin \theta < S^z_{2}>\label{aterm}\\
B_n &=& 4J^{m} < S^z_{n}> \gamma (\cos\theta-1)
-4D \sin \theta < S^z_{n}>\gamma,\ \ \  n=1,2\label{bterm}\\
C_n &=& 2J^{m} < S^z_{n}>, \ \ \  n=1,2 \label{cterm}
\end{eqnarray}
Note that by symmetry, one has $<S^z_{1}>=<S^z_{2}>$.

\begin{figure}[ht!]
\centering
\includegraphics[width=7cm,angle=0]{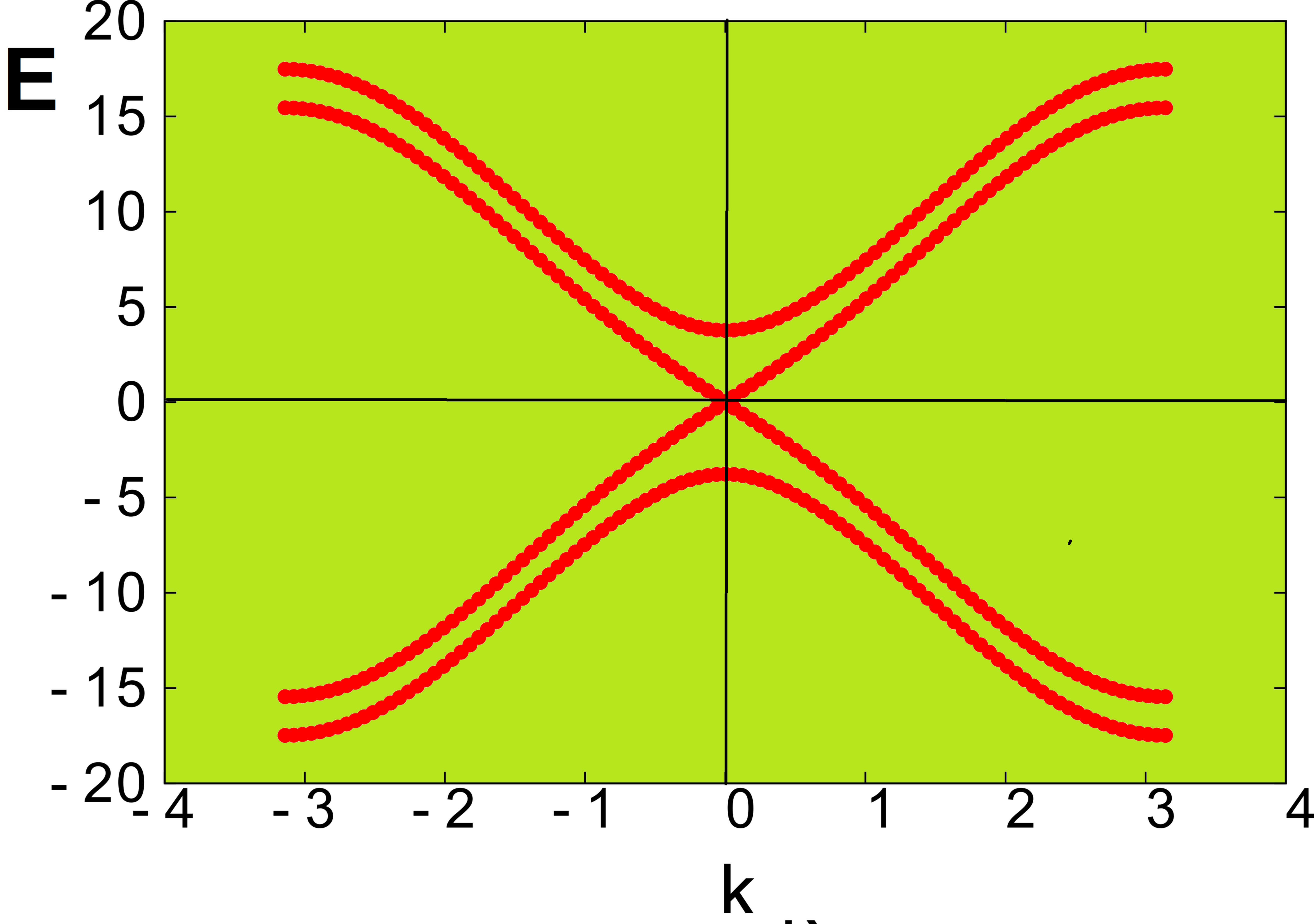}
\caption{Spin-wave energy $E$ versus $k=k_x=k_y$ at $T=0$ for a bilayer with $\theta=0.6$ radian.
\label{SWbilayer}}
\end{figure}

We show in Fig. \ref{SWbilayer} the SW spectrum of the bilayer case for a strong value $\theta=0.6$ radian. There are two important points:

(i) the first mode has the $E\propto k$ antiferromagnetic behavior at the long wave-length limit for this strong $\theta$,

(ii) the higher mode has $E\propto k^2$ which is the ferromagnetic wave due to the parallel NN spins in the $z$ direction.

%Fig13

In conclusion of this section, we  emphasize that the DM interaction affects strongly the SW mode at $k \rightarrow 0$. Quantum fluctuations in competition with thermal effects cause the cross-over of magnetizations of different $\theta$: in general stronger $\theta$ yields stronger spin contraction at and near $T=0$ so that the corresponding spin length is shorter. However at higher $T$, stronger $\theta$ means stronger $J^{mf}$ which yields stronger magnetization. It explains the cross-over at moderate $T$.

\section{Monte Carlo results}\label{MC}

We have used the Metropolis algorithm \cite{Landau09,Brooks11} to calculate physical quantities of the system at finite temperatures $T$. As said above, we use mostly the size $N\times N \times L$ with $N=40$ and thickness $L=L_m+L_f=8$ (4 magnetic layers, 4 ferroelectric layers).  Simulation times are $10^5$ Monte Carlo steps (MCS) per spin for equilibrating the system and $10^5$ MCS/spin for averaging.  We calculate the internal energy and the layer order parameters of the magnetic ($M_m$) and ferroelectric ($M_f$) films.

The order parameter $M_f(n)$ of layer $n$ is defined as
\begin{equation}\label{eq-orpar1}
M_f(n)=\frac{1}{N^{2}}\langle{|\sum_{i\in n} P_{i}^z|}\rangle
\end{equation}
where $ \langle{...}\rangle$ denotes the time average.

The definition of an order parameter for a skyrmion crystal is not obvious. Taking advantage of the fact that we know the GS, we define the order parameter as the projection of an actual spin configuration at a given $T$ on its GS and we take the time average.  This order parameter of layer $n$ is thus defined as
\begin{equation}\label{OP}
M_m(n)=\frac{1}{N^2(t_a-t_0)}\sum_{i\in n} |\sum_{t=t_0}^{t_a} \mathbf S_i (T,t)\cdot \mathbf S_i^0(T=0)|
\end{equation}
where $\mathbf S_i (T,t)$ is the $i$-th spin at the time $t$, at temperature $T$, and $\mathbf S_i (T=0)$ is its state in the GS. The order parameter $M_m(n)$ is close to 1 at very low $T$ where each spin is only weakly deviated from its state in the GS. $M_m(n)$  is zero when every spin strongly fluctuates in the paramagnetic state.
The above definition of $M_m(n)$  is similar to the Edward-Anderson order parameter used to measure the degree of freezing in spin glasses \cite{Mezard}: we follow each spin with time evolving and take the spatial average at the end.
The total order parameters $M_m$ and $M_f$ are the sum of the layer order parameters, namely $M_m=\sum_n M_m(n)$ and $M_f=\sum_n M_f(n)$.

\begin{figure}[h]
\vspace{10pt}
\begin{center}
\includegraphics[scale=0.37]{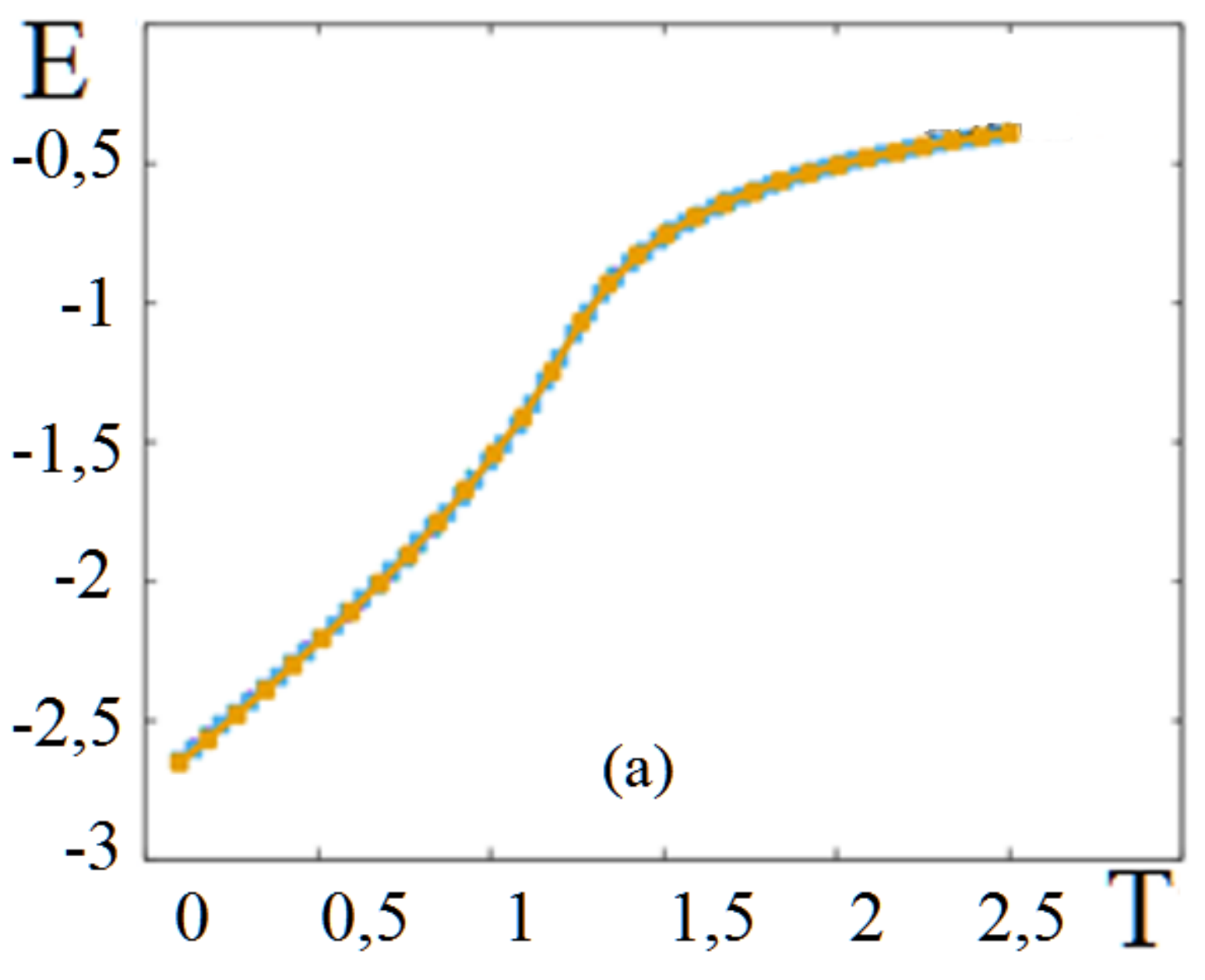}
\includegraphics[scale=0.37]{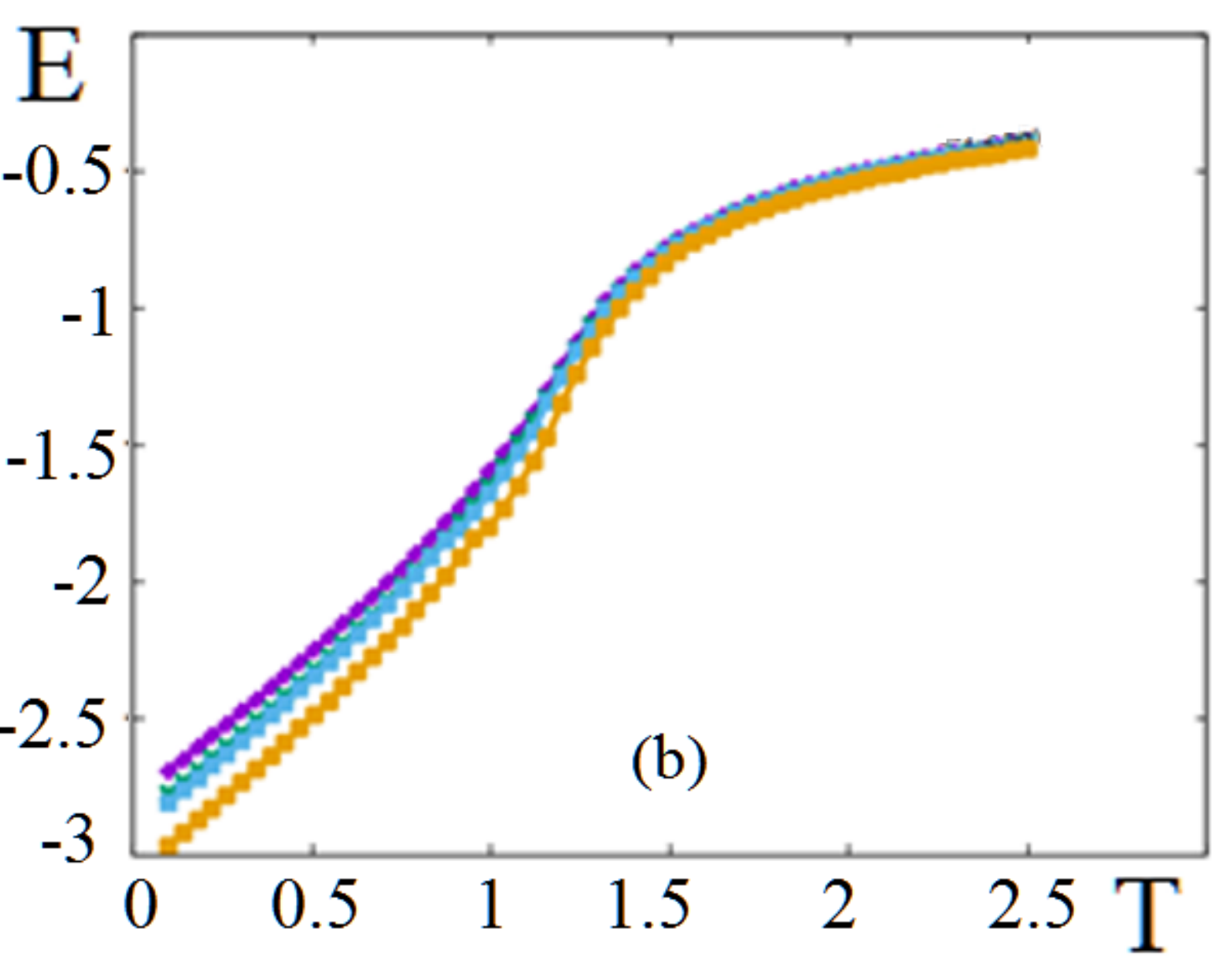}
\end{center}
\vspace{10pt} \caption{Energy of the magnetic film versus temperature $T$ for (a) $J^{mf}=-0.1, J^{mf}=-0.125, J^{mf}=-0.15, J^{mf}=-0.2$  (all the lines are the same, see text for comments); (b)  $J^{mf}=-0.45$ (purple line), $J^{mf}=-0.75$ (green line), $J^{mf}=-0.85$ (blue line) and  $J^{mf}=-1.2$ (gold line),  without an external magnetic field.} \label{fig6} \vspace{10pt}
\end{figure}

In Fig.\ref{fig6} we show the dependence of energy of the magnetic film
versus temperature, without an external magnetic field,
 for various values of the interface
magnetoelectric interaction: in Fig.\ref{fig6}a for weak values $J^{mf}=-0.1, J^{mf}=-0.125,
J^{mf}=-0.15, J^{mf}=-0.2$, and in Fig.\ref{fig6}b for stronger values $J^{mf}=-0.45,J^{mf}=-0.75,
J^{mf}=-0.85, J^{mf}=-1.2$.

%Fig14

As said in the GS determination, when $J^{mf}$ is weak, the GS is composed with large ferromagnetic domains at the interface (see Fig. \ref{fig4}). Interior layers are still ferromagnetic.  The energy is therefore does not vary with weak values of $J^{mf}$ as seen in Fig. \ref{6}a.  The phase transition occurs at the curvature change, namely maximum of the derivative or maximum of the specific heat, $T_c^m\simeq 1.25$.
Note that the energy at $T=0$ is equal to -2.75  by extrapolating the curves in Fig. \ref{fig6}a to $T=0$. This value is just the sum of energies of the spins across the layers: 2 interior spins with 6 NN, 2 interface spins with 2 NN. The energy per spin is thus (in ferromagnetic state): $E=-(2\times 6+ 2\times 5)/(4\times 2)=-2.75$ (the factor 2 in the denominator is to remove the bond double counting in a crystal).

For stronger values of $J^{mf}$, the curves shown in Fig. \ref{fig6}b indicate a deviation of the ferromagnetic state due to the non collinear interface structure.  Nevertheless, we observe the magnetic transition at almost the same temperature, namely $T_c^m\simeq 1.25$. It means that spins in interior layers dominate the ordering.

\begin{figure}[h]
\vspace{10pt}
\begin{center}
\includegraphics[scale=0.37]{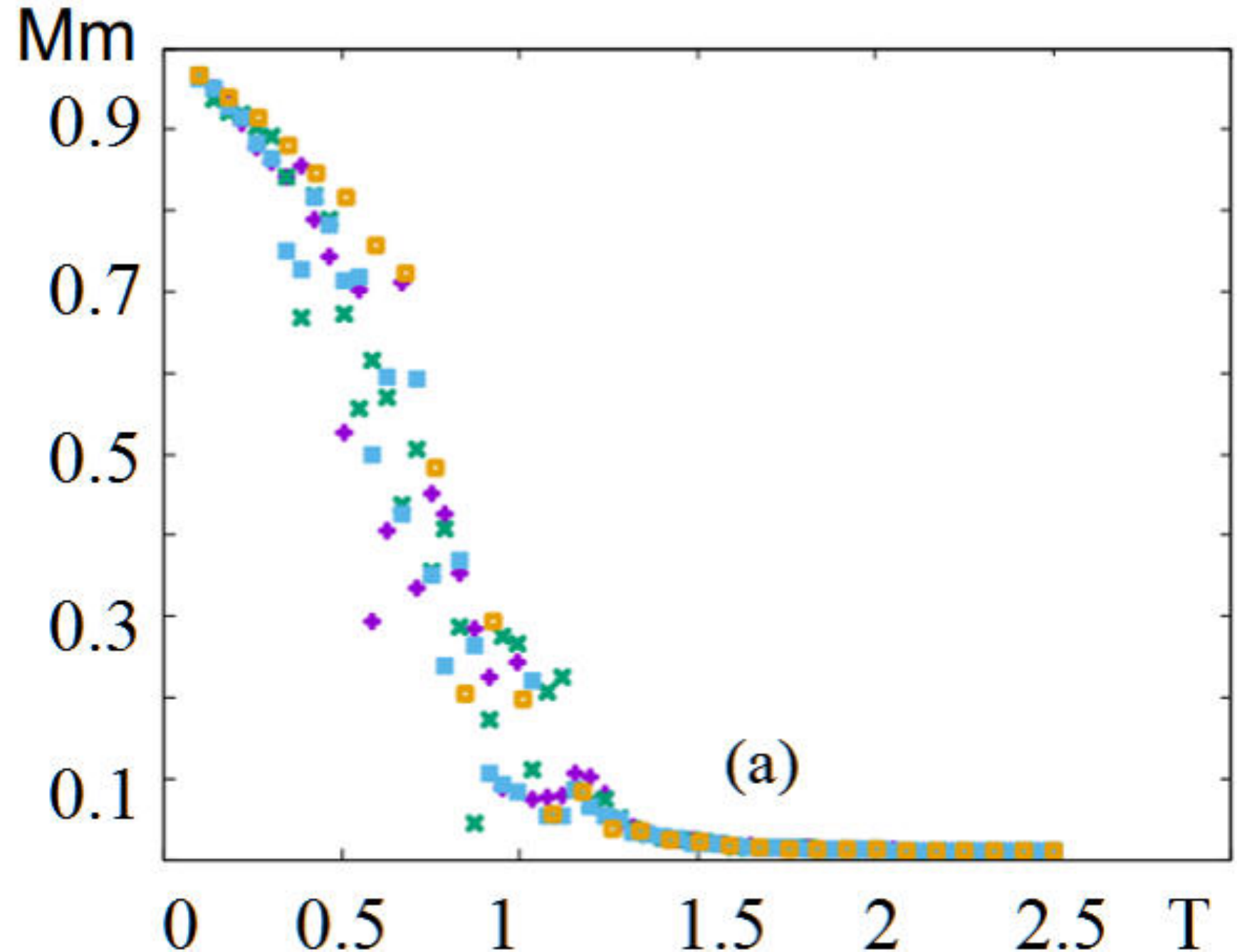}
\includegraphics[scale=0.07]{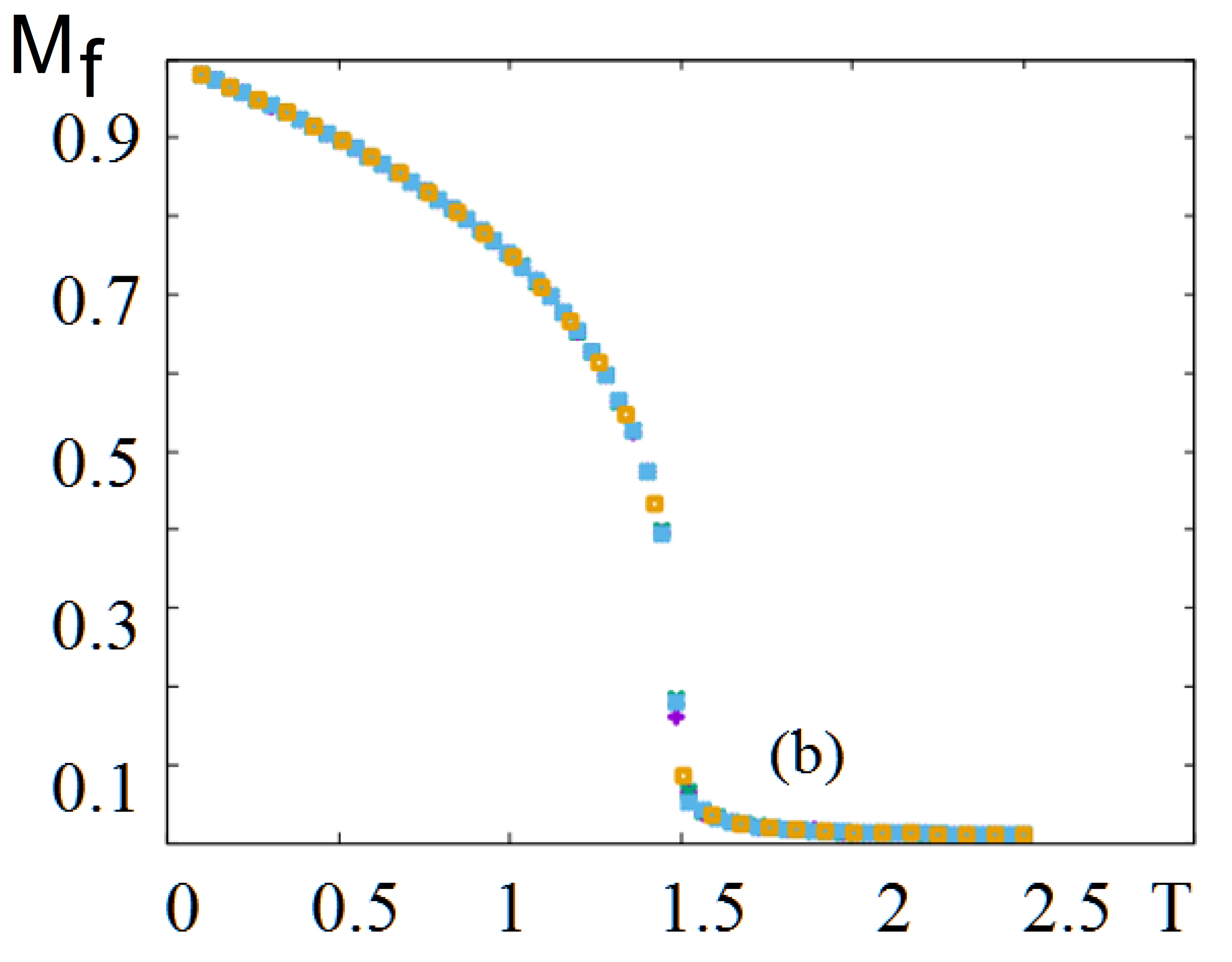}
\end{center}
\vspace{10pt} \caption{(a) Order parameter of the magnetic film $M_m$ versus $T$; (b) Order parameter of the ferroelectric film $M_f$ versus $T$,  for $J^{mf}=-0.1$ (purple dots), $J^{mf}=-0.125$ (green dots), $J^{mf}=-0.15$ (blue dots), $J^{mf}=-0.2$ (gold dots),  without an external magnetic field.} \label{fig7} \vspace{10pt}
\end{figure}

We show in Fig. \ref{fig7}  the total order parameters of
the magnetic film $M_m$  and the ferroelectric film $M_f$ versus $T$
for various values of the parameter of the magnetoelectric
interaction  $J^{mf}=-0.1, -0.125, -0.15, -0.2$
and for  $J^{mf}=-0.45,-0.75, -0.85, -1.2$, without an external magnetic field.  Several remarks are in order:

i) For the magnetic film, $M_m$ shows strong fluctuations but we still see that all curves fall to zero at $T_c^m\simeq 1.25$. These fluctuations come from non uniform spin configurations and also from the nature of the Heisenberg spins in low dimensions \cite{Mermin}.

ii) For the ferroelectric film, $M_f$ behaves very well with no fluctuations. This is due to the Ising nature of electric polarizations supposed in the present model.  The ferroelectric film undergoes a phase transition at $T_c^f\simeq 1.50$.

iii) There are thus two transitions, one magnetic and one ferroelectric, separately.

%Fig15

\begin{figure}[h]
\vspace{10pt}
\begin{center}
\includegraphics[scale=0.085]{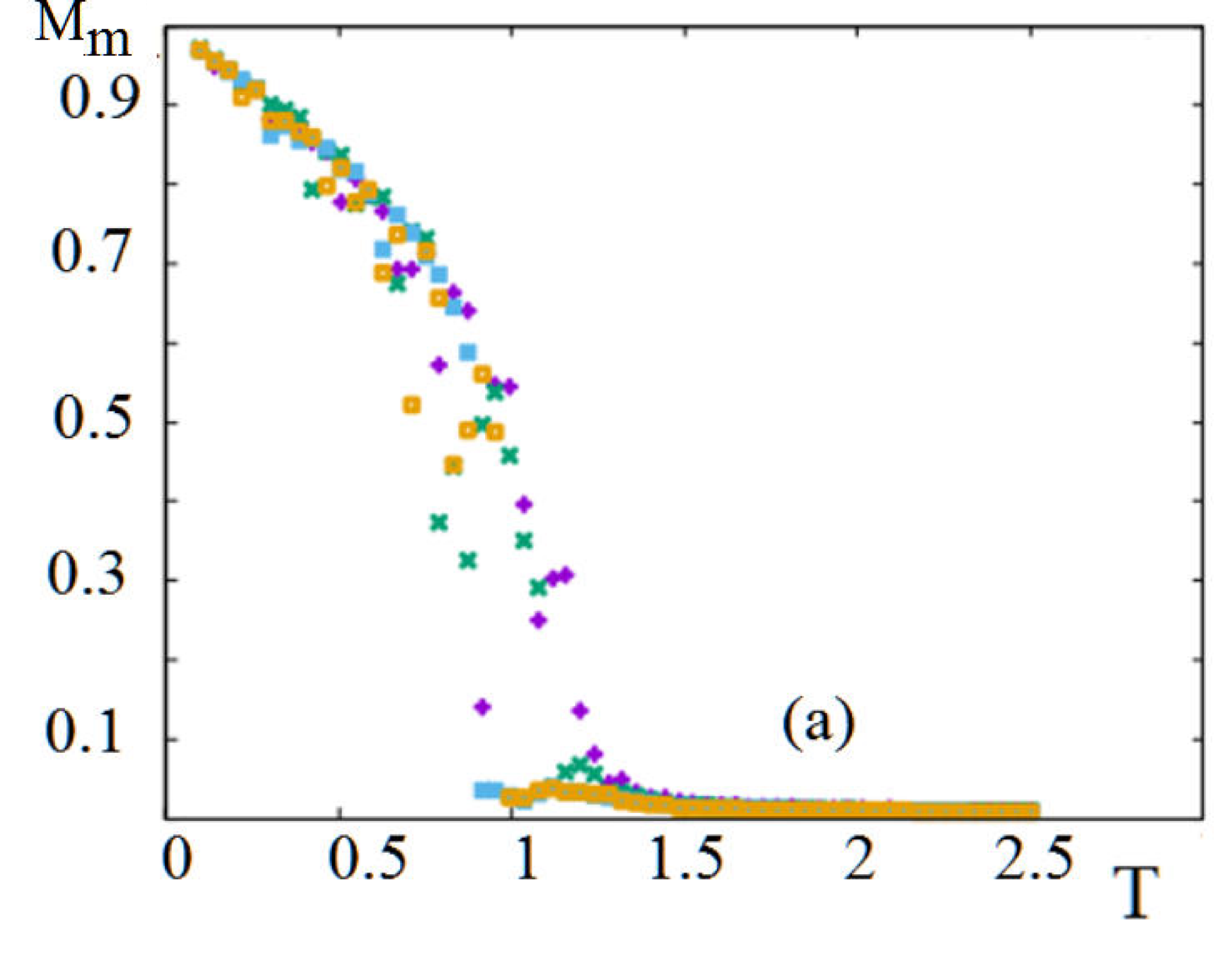}
\includegraphics[scale=0.37]{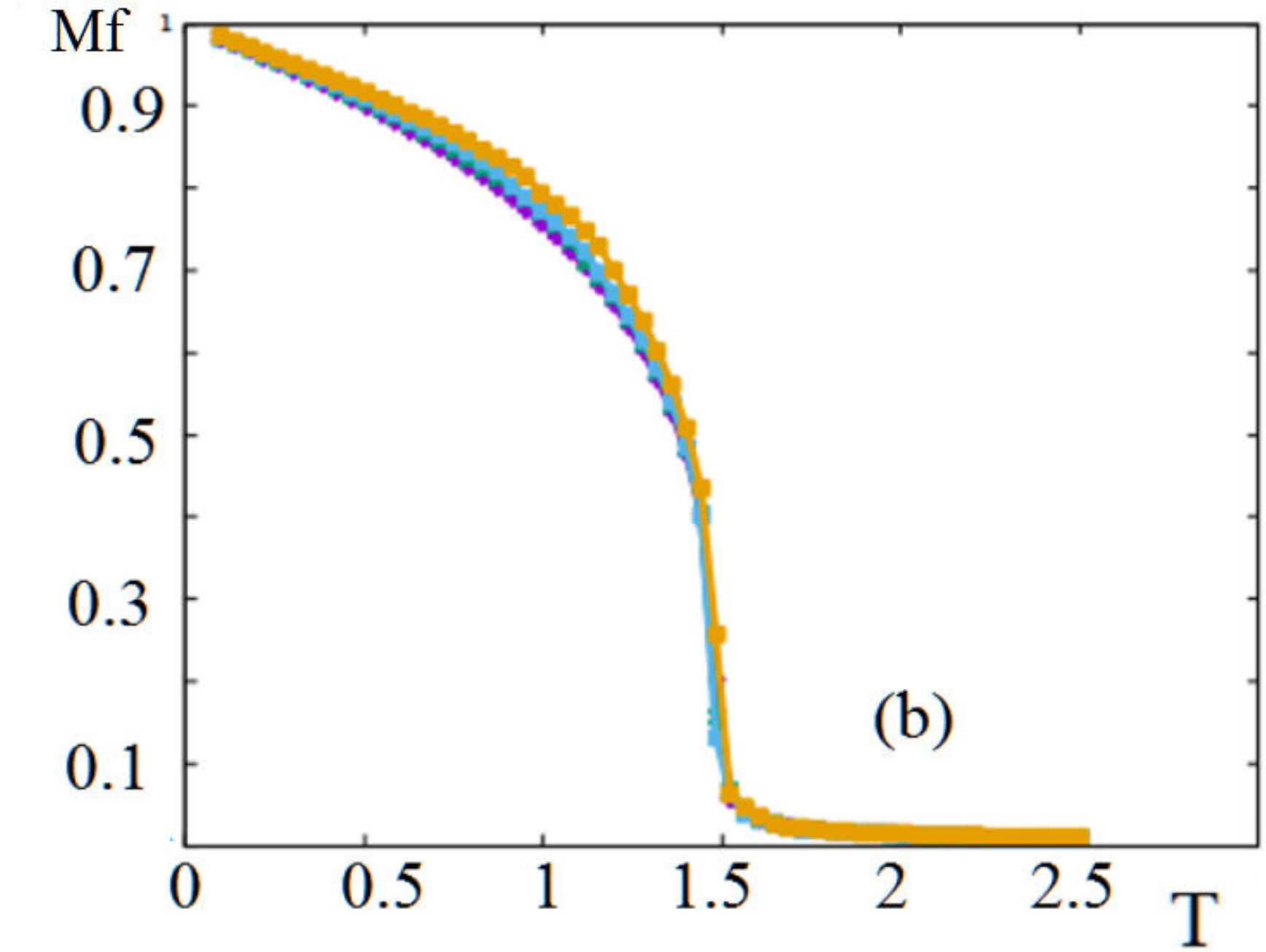}
\end{center}
\vspace{10pt} \caption{(a) Order parameter of the magnetic film versus $T$; (b) Order parameter of the ferroelectric film  versus $T$ for $J^{mf}=-0.45$ (purple dots), $J^{mf}=-0.75$ (green dots), $J^{mf}=-0.85$ (blue dots)
and $J^{mf}=-1.2$ (gold dots), without an external magnetic field.} \label{fig8} \vspace{10pt}
\end{figure}

We show in Fig. \ref{fig8} the order parameters of the magnetic and ferroelectric films at strong values of $J^{mf}$ as functions of $T$, in zero field.  We observe that the stronger $J^{mf}$ is, the lower $T_c^m$ becomes.The ferroelectric $T_c^f$ does not change as expected.

%Fig16

\begin{figure}[h]
\vspace{10pt}
\begin{center}
\includegraphics[scale=0.07]{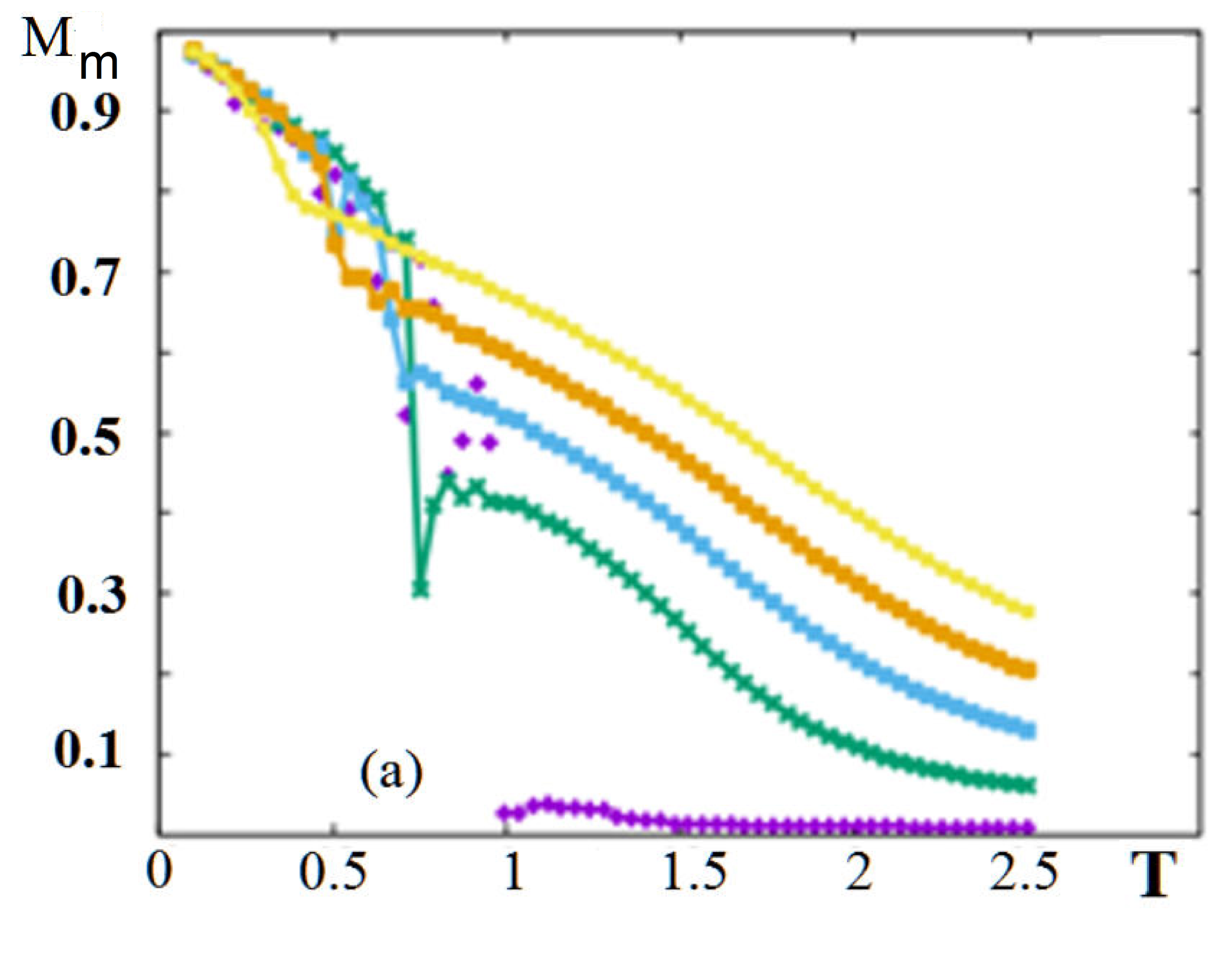}
\includegraphics[scale=0.07]{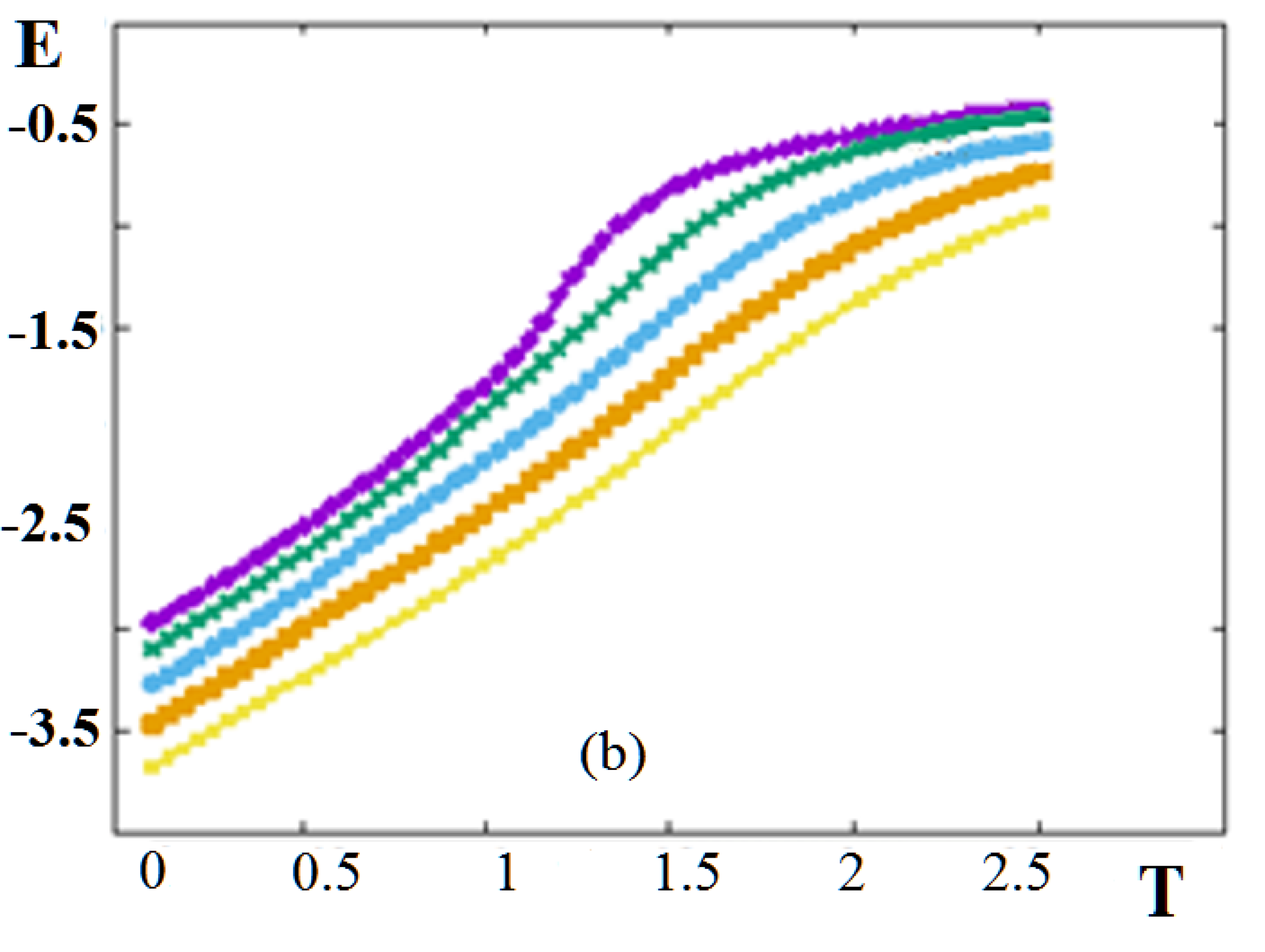}
\end{center}
\vspace{10pt} \caption{(a) Temperature dependence of (a) the magnetic order parameter; (b) the magnetic energy for $H=0$ (purple dots), $H=0.25$ (green line), $H=0.5$ (blue line), $H=0.75$ (gold line),
  $H=1$ (yellow line). The interface magnetoelectric interaction is $J^{mf}=-1.2$.} \label{fig9} \vspace{10pt}
\end{figure}

We examine the field effects now. Figure \ref{fig9}  shows the order parameter
and the energy of the magnetic film versus $T$, for
various values of the external magnetic field. The interface magnetoelectric interaction is $J^{mf}=-1.2$. Depending on the magnetic field, the non collinear spin configuration survives up to a temperature between 0.5 and 1 (for $H=0$). After the transition, spins align themselves in the field direction, giving a large value of the order parameter (Fig. \ref{fig9}a). The energy shows a sharp curvature change only for $H=0$, meaning that the specific heat is broadened more and more with increasing $H$.

%Fig17

We consider now the case of very strong interface couplings.

\begin{figure}[h]
\vspace{10pt}
\begin{center}
\includegraphics[scale=0.37]{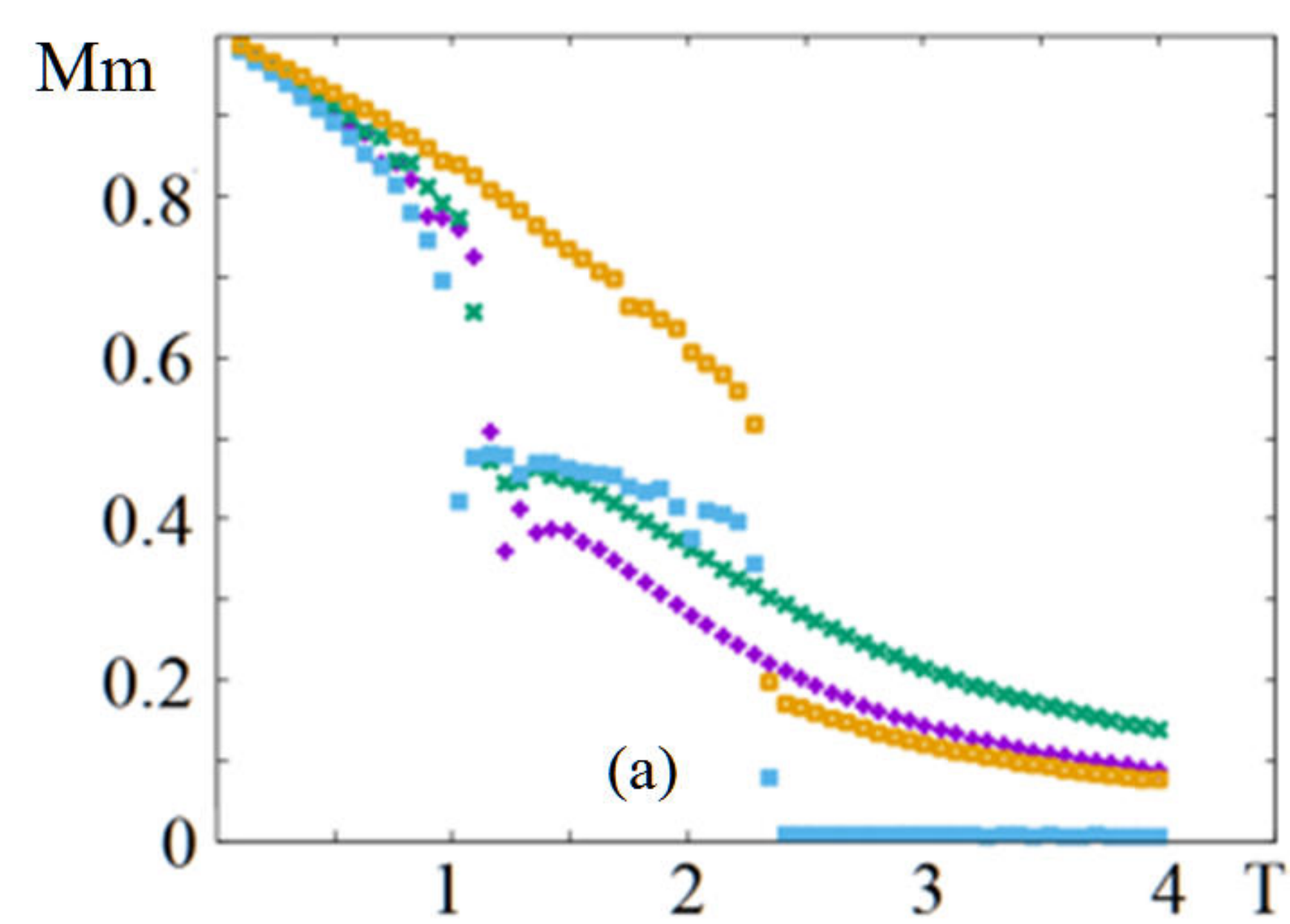}
\includegraphics[scale=0.35]{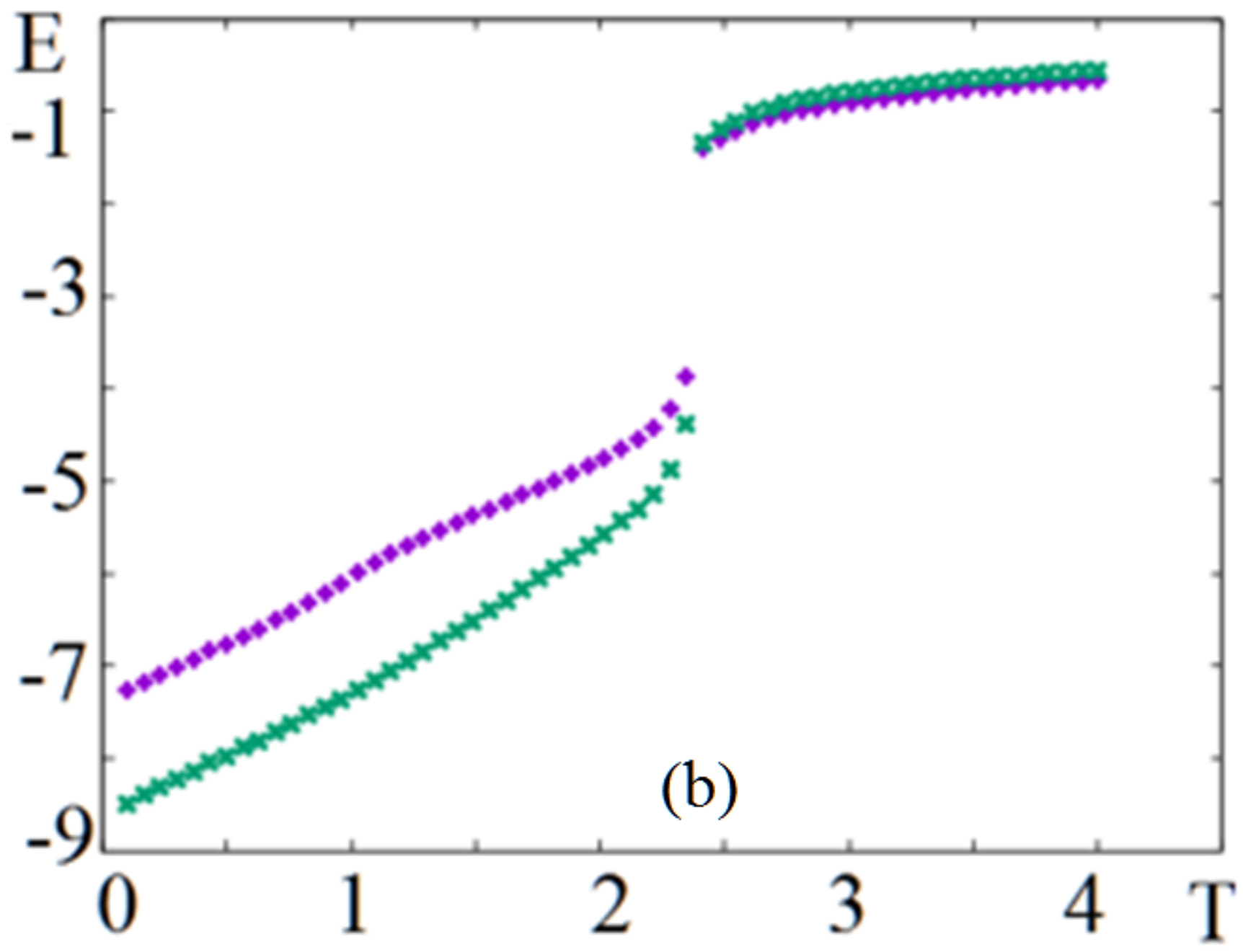}
\end{center}
\vspace{10pt} \caption{(a) Order parameter of magnetic
film versus $T$. The purple and green dots correspond to
$M$ for $(J^{mf}= -2.5, H^z=1)$ and $(J^{mf}= -2.5, H^z=1.5$),
blue and gold dots correspond to  $M$ for
 $(J^{mf}= -6, H^z=1)$ and $(J^{mf}= -6, H^z=0)$. (b) Energies of magnetic (purple
dots) and ferroelectric (green dots) subsystems versus $T$ for
$(J^{mf}= -6,H=0)$.} \label{fig13} \vspace{10pt}
\end{figure}

Figure \ref{fig13}a shows the magnetic order
parameter versus $T$. The purple and green lines correspond to
$M$ for $J^mf= -2.5$ with $H^z=1.0$ and $H^z=1.5$, respectively;
the blue and gold lines correspond to  $M$ for  $J^{mf}= -6$
with $H^z=0$ and $H^z=1$.
These curves indicate first-order phase
transitions
at $T_c^m=1.05$ for $(J^{mf}= -2.5,H^z=1)$ (purple),
at $T_{c}^m=1.12$ for $(J^{mf}= -2.5, H^z=1.5$) (green)
and at $T_{c}^m=2.25$ for $(J^{mf}= -6, H^z=1)$ (gold).
In the case of zero field, namely
$(J^{mf}=-6, H^z=0)$ (blue),  one has two first-order phase transitions occurring at
$T_{c1}=1.05 $ and $T_{c2}=2.19$.

Figure \ref{fig13}b shows the magnetic (purple) and ferroelectric (green) energies versus $T$ for $(J^{mf}=-6, H^z=0)$. One sees the discontinuities of these curves at $T_c \simeq 2.29$, indicating the first-order transitions for both magnetic and ferroelectric at the same temperature.  In fact, with such a strong $J^{mf}$ the transitions in both magnetic and ferroelectric films are driven by the interface, this explains the same $T_c$ for both.

%Fig18

\begin{figure}[h]
\vspace{10pt}
\begin{center}
\includegraphics[scale=0.07]{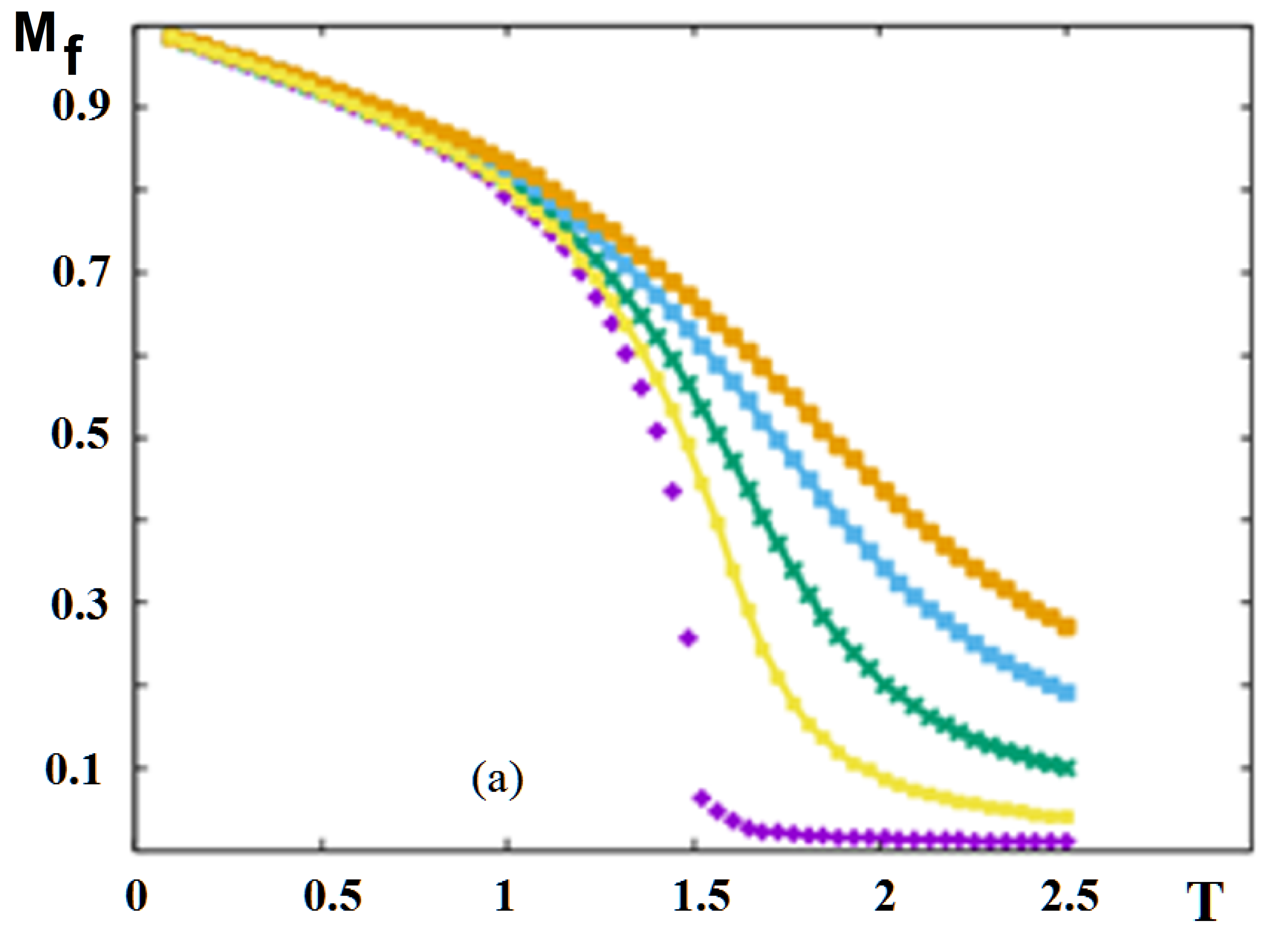}
\includegraphics[scale=0.30]{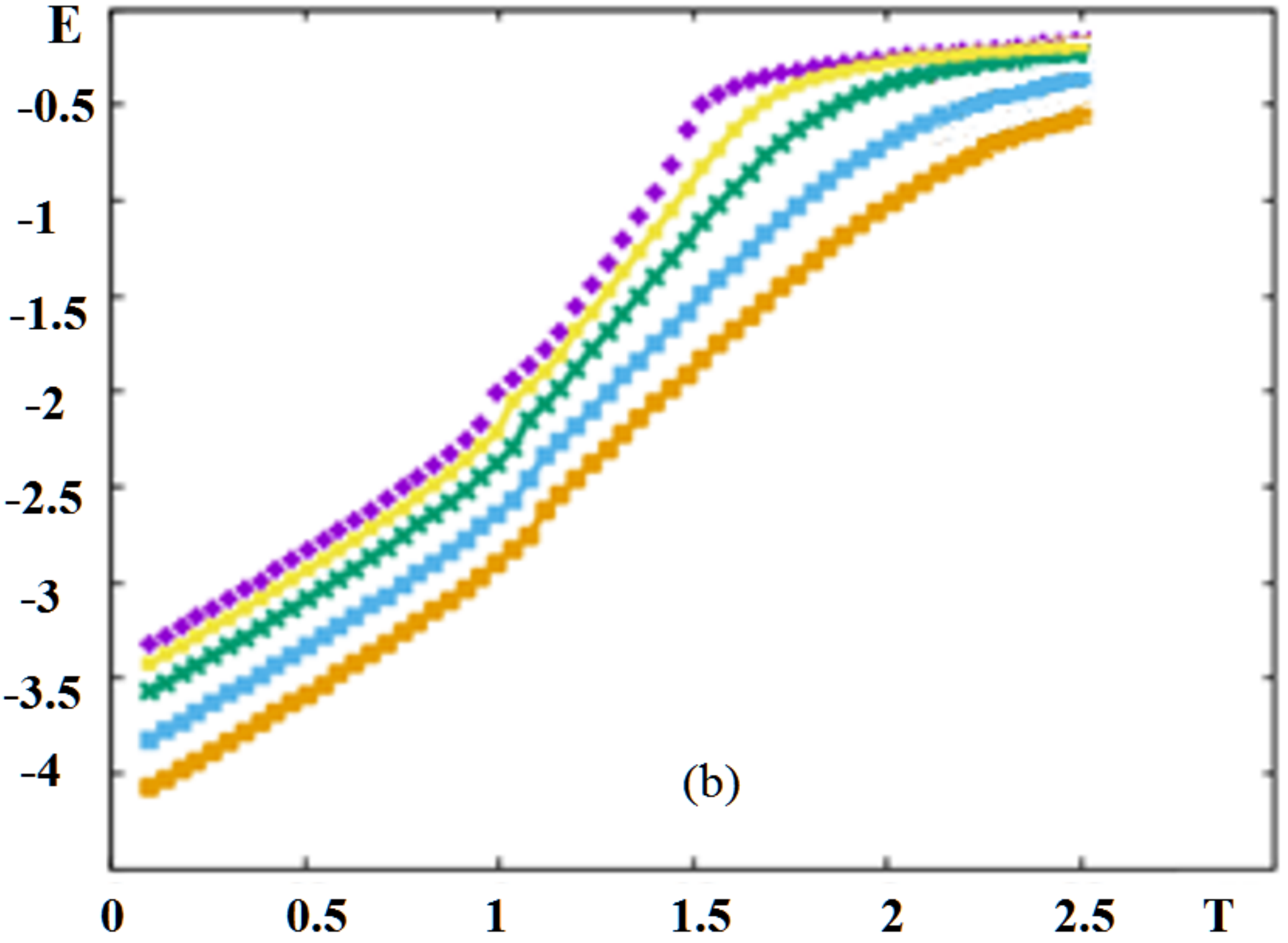}
\end{center}
\vspace{10pt} \caption{(a) Order parameter and (b) energy of ferroelectric film, versus temperature
for $E=0$ (purple dots), $E=0.25$ (green line), $E=0.5$ (blue line), $E=0.75$ (gold line), $E=1$ (yellow line).
The interface magnetoelectric interaction is $J^{mf}=-1.2$ } \label{fig10} \vspace{10pt}
\end{figure}

Let us show the effect of an applied electric field. For the ferroelectric film, polarizations are along the $z$ axis so that an applied electric field $\mathbf E$ along this direction will remove the phase transition: the order parameter never vanishes when $E\neq 0$. This is seen in Fig. \ref{fig10}.  Note that the energy has a sharp change of curvature for $E=0$ indicating a transition, other energy curves with $E\neq 0$ do not show a transition. One notices some anomalies at $T\sim 1-1.1$ which are due to the effect of the magnetic transition in this temperature range.

%Fig19

%In Fig. \ref{fig11} $(a,b)$ shown the dependence of surface (a) and interior
%(b) layers magnetizations  versus temperature for various values of
%strength of the electric field. Parameter of the magnetoelectric
%interaction $J^{mf}=-1.2$.
%%Fig16
%\begin{figure}[h]
%\vspace{10pt}
%\begin{center}
%\includegraphics[scale=0.33]{Fig 11a.pdf}
%\includegraphics[scale=0.33]{Fig 11b.pdf}
%\end{center}
%\vspace{10pt} \caption{The dependence of surface layer magnetization (a) and interior layer magnetization (b)
% versus temperature for $H=0$ (purple dots), $H=0.25$ (green line), $H=0.5$ (blue line), $H=0.75$ (gold line),
% $H=1$ (yellow line). Parameter of the magnetoelectric interaction $J^{mf}=-1.2$.  } \label{fig11} \vspace{10pt}
%\end{figure}

%Fig. \ref{fig12} show the dependence surface polarization versus
%temperature for various values of strength of external the electric
%field. Parameter of the magnetoelectric interaction $J^{mf}=-1.2$.
%%Fig17
%\begin{figure}[h]
%\vspace{10pt}
%\begin{center}
%\includegraphics[scale=0.40]{Fig 12.pdf}
%\end{center}
%\vspace{10pt} \caption{The dependence surface polarization versus temperature for $E=0$ (purple dots),
%$E=0.2$5 (green line), $E=0.5$ (blue line), $E=0.75$ (gold line), $E=1$ (yellow line).
%Parameter of the magnetoelectric interaction $J^{mf}=-1.2$.} \label{fig12} \vspace{10pt}
%\end{figure}

\section{Conclusion}\label{Concl}

We have studied in this paper a new model for the interface coupling between a magnetic film and a ferroelectric film in a superlattice. This coupling has the form of a Dzyaloshinskii-Moriya (DM) interaction between a polarization and the spins at the interface.

The ground state shows uniform non collinear spin configurations in zero field and skyrmions in an applied magnetic field.
We have studied spin-wave (SW) excitations in a monolayer and in a bilayer in zero field by the Green's function method. We have shown the strong effect of the DM coupling on the SW spectrum as well as on the magnetization at low temperatures.

Monte Carlo simulation has been used to study the phase transition occurring in the superlattice with and without applied field.  Skyrmions have been shown to be stable at finite temperatures. We have also shown that the nature of the phase transition can be of second or first order, depending on the DM interface coupling.

The existence of skyrmions confined at the magneto-ferroelectric interface is very interesting.  We believe that it can be used in transport applications in spintronic devices. A number of applications using skyrmions has been already mentioned in the Introduction.

\section*{Acknowledgment}
One of us (IFS) wishes to thank Campus France for a financial support (contract P678172A) during the course of the present work.

\section*{References}
\label{sect-ref}
\bibliographystyle{elsarticle-num}
\bibliography{bibliography}
\end{document}